\documentclass[
aip,
dcolumn,
bm,
nofootinbib,
notitlepage]{revtex4-2}
\usepackage{graphicx}
\usepackage{color}
\usepackage[latin1]{inputenc}
\usepackage[T1]{fontenc}
\usepackage{amsmath}
\usepackage{amssymb}
\usepackage{braket}
\usepackage{mathrsfs}
\usepackage{hyperref}
\usepackage{natbib} 
\usepackage{listings}
\usepackage{tikz}
\usepackage{tcolorbox}
\usetikzlibrary{mindmap}

\allowdisplaybreaks

\usepackage{flafter}
\usepackage{float}

\definecolor{lightblue}{rgb}{0.145,0.6666,1}
\begin{document}
	\title{High-harmonic generation under electronic strong coupling: A 
		time-dependent combined quantum electrodynamics / quantum chemistry  study}
	
	\author{Paul A. Albrecht}
	\affiliation{Institut f\"ur Chemie, Universit\"at Potsdam, Karl-Liebknecht-Stra\ss{}e 24-25, D-14476 Potsdam-Golm, Germany}
	
	\author{Eric W. Fischer}
	\affiliation{Institut f\"ur Chemie, Humboldt-Universit\"at zu Berlin, Brook-Taylor-Stra\ss{}e 2, D-12489, Berlin, Germany}
	
	\author{Tillmann Klamroth}
	\affiliation{Institut f\"ur Chemie, Universit\"at Potsdam, Karl-Liebknecht-Stra\ss{}e 24-25, D-14476 Potsdam-Golm, Germany}

	\author{Peter Saalfrank}
	\email{peter.saalfrank@uni-potsdam.de}
	\affiliation{Institut f\"ur Chemie, Universit\"at Potsdam, Karl-Liebknecht-Stra\ss{}e 24-25, D-14476 Potsdam-Golm, Germany}
	\affiliation{Institut f\"ur Physik und Astronomie, Universit\"at Potsdam, Karl-Liebknecht-Stra\ss e 24-25, D-14476 Potsdam-Golm, Germany}
	
	\date{\today}
	
	\let\newpage\relax
	
	\begin{abstract}
The creation of light-matter hybrid states, polaritons, in a cavity offers new intriguing opportunities to manipulate the electronic structure and electron dynamics of atoms and molecules.
Here, we investigate the effect of electronic strong coupling (ESC) between atoms or molecules and field modes of a Fabry-P\'erot cavity on High-Harmonic Generation (HHG) spectra within a theoretical model study.
We assume that the atom or molecule 
is driven by an intense classical laser field, giving rise 
to HHG, while being strongly coupled to quantized cavity modes as described by the Pauli-Fierz Hamiltonian in the framework of molecular quantum electrodynamics (QED).
Specifically, as a test case, 
we first consider a model Hamiltonian of a one-dimensional 
hydrogen atom coupled to a 
cavity mode, which can be treated ``numerically exact'' 
using grid methods. Further, a hydrogen molecule coupled to a cavity 
mode is considered and treated 
within a recently suggested QED-TD-CI (Quantum Electrodynamics Time-Dependent Configuration 
Interaction) method [Weidman {\em et al.}, J. Chem. Phys. {\bf 160}, 094111 (2024)]. 
The resulting HHG spectra show (i) 
 a suppression of the harmonic cutoff in line with 
excitation of quantum light in the cavity and, in some cases, (ii) enhancement of some harmonics 
 of the coupled light-matter system.
	\end{abstract}
	
	\let\newpage\relax
	\maketitle
	\section{Introduction}
	\label{sec.intro}
	Last decades' advances in the field of attosecond dynamics, acknowledged with the 2023 Nobel prize in physics, enable the investigation of electron dynamics on their natural timescale. \cite{Corkum2007,Sansone2010,Nisolini2017,Gaumnitz2017}
	A central process in this regard is High-Harmonic Generation (HHG). This process is the source of coherent, high-energy frequency combs, used for the creation of ultra-short pulses of light to investigate the attosecond domain.
	Additionally, HHG is intimately related to electron dynamics, usually described by Corkum's three-step model:
	The creation of an excited electron, moving away from the nucleus, so first being born and second accelerated by the incoming laser pulse, and third, recombining at the parent ion after some time.\cite{Corkum1993,Corkum2007}
	This process results in non-linear scattering of light into multiple integer values of the incoming laser photon energy. In the context of attosecond pulse 
		generation by HHG light, the ``cutoff'' of the HHG spectra, {\em i.e.}, the frequency 
		$\omega_\mathrm{cut}$ beyond which 
		the intensity of the HHG spectrum falls off exponentially, is of central importance. According to Corkum's model, 
		$\hbar \omega_\mathrm{cut} = 3.17 U_p + I_p$, {\em i.e.}, the cutoff
		depends on pulse properties, contained in the ponderomotive energy, $U_p$, which is 
 proportional to 
 the laser intensity, $I$, 
		and on properties 
		of the atomic / molecular system, {\em e.g.}, the ionization potential, 
		$I_p$. 
		\\
		
		The question arises as to whether further ``control'' strategies and 
		associated parameters are possible by which HHG cutoffs and HHG responses in general, could be affected. A possibility could be those provided 
		by the (strong) coupling of the atom's / molecule's electrons to the quantized electromagnetic 
		field in a cavity. Shining light on this possibility is a goal 
		of this paper. 
		Another goal is to test a recently proposed method for treating 
		many-electron dynamics in cavities, called the QED-TD-CI (Quantum Electrodynamics Time-Dependent Configuration
		Interaction) method,\cite{qedtdci} now 
		for the case of laser-driven, large-amplitude  motion of electrons 
		as it occurs in HHG. 
	\\ 
	
	Experimentally, ``cavity-enhanced'' High-Harmonic Generation was established 
	a long time 
		ago as a key to 
		attosecond pulse generation, by using arrays of mirrors
	to recycle the generating pulse passing through the medium and enabling the generation of high energy, high repetition rate HHG pulses.\cite{Jour2005,Jones2005,Hogner2019} In this scenario, however, no quantum 
		light is involved whatsoever and high-intensity laser pulses are used which can well be described by 
		classical electrodynamics.
	Another way to enhance HHG is to generate pulses in the vicinity of surface polaritons (plasmons) on nanostructures, creating a coupling between the incident light and the plasmon.\cite{Kim2008,Park2011,Ebadian2017}
		This has also been attempted to model theoretically, with classical field and light interaction during the HHG process.\cite{Shahnavaz2021}
	Recently, outside the context of cavity or plasmon physics, however,
		the role of field quantization of the {driving laser} on HHG has been examined 
		theoretically, with the result that quantum light can produce 
		HHG cutoffs that depend nonlinearly on $I$ and can be higher than their classical-light analogs.\cite{Gorlach2023}
		\\

		In what follows, we shall 
		consider the possibility to manipulate HHG 
		by electronic strong coupling to the quantized cavity field. We are interested in scenarios where an exciting,  
		intense laser pulse can still be described by classical electrodynamics, but 
		the process is, perhaps, affected by the presence of the quantized electromagnetic field of a cavity, which 
		hosts the atom / molecule. 
 Moreover, our work extends recent results\cite{Aklilu2024} on HHG generation under ESC for a one-electron system to the many-electron scenario.
	\\

	While single-electron / single-cavity mode models can be treated 
 numerically more or less
		``exact'' (see below), 
		the description of 
		(strong) light-matter coupling between quantized cavity modes and atomic / molecular many-electron systems 
		require more approximate (electronic structure) methods.
	Recently, {\em ab initio} electronic structure theory
  for realistic molecules 
		was extended to the QED frame, {\em e.g.}, QED Hartree-Fock (QED-HF),\cite{Haugland2020,cavityionization},
		QED Density Functional Theory (QED-DFT),\cite{Tokatly2013,Ruggenthaler2014,Flick2017atoms}
		QED Coupled Cluster theory (QED-CC),\cite{Haugland2020,cavityionization} and 
		even QED Full Configuration interaction (QED-FCI)\cite{Haugland2020,Haugland2021} were developed. 
In this work, we exploit a truncated QED-CI approach recently 
		proposed in a {time-dependent} 
		context, 
		the QED-TD-CI method,\cite{qedtdci} which is particularly well-suited for electronic polariton dynamics in the ESC regime.
	%
	Specifically, the QED-TD-CI singles (QED-TD-CIS) method will be employed to study the HHG spectrum of a laser-driven H$_2$ molecule in a cavity.
	To account for ionization, we extend ionization 
 models developed earlier in the context of TD-CIS\cite{bedurke2019discriminating,saalfrank2020molecular,bedurke2021many}, to the ESC scenario.
	 We assume here that 
		the  incoming, classical laser field couples to the molecule only, 
		not directly to the cavity, but the molecule 
		couples to the quantized cavity modes.
 Possible ionization losses are accounted for.
	To support the molecular calculations based on  QED-TD-CI, 
	we also provide ``numerically exact'' 
	solutions for HHG spectra of a one-dimensional single-electron atomic model
	coupled to a harmonic cavity mode.
	\\
	
	The paper is structured as follows. In Sec.\ref{sec.theory}, we introduce the Pauli-Fierz Hamiltonian for ESC, the resulting time-dependent
		Schr\"odinger equation (TDSE) as well as methods to calculate HHG spectra. Numerical approaches towards the solution of the TDSE for the one-dimensional atom model (by grid methods) and for the H$_2$ molecule (by QED-TD-CIS) will be described in Sec.\ref{sec3}. We present and discuss our results in Sec.\ref{sec.results}. Finally, Sec.\ref{sec.conclusion} concludes this work.
	In what follows, we use atomic units throughout but 
		indicate $\hbar$ explicitly.
	\section{Theory}
	\label{sec.theory}
	\subsection{The Polaritonic Pauli-Fierz Hamiltonian}
	We consider a single atom or molecule under electronic strong coupling with a single effective transverse field mode of a Fabry-P\'erot cavity as described by the polaritonic Pauli-Fierz Hamiltonian\cite{Flick2017,Flick2017atoms,Schaefer2018,Li2021,Fischer2023cbo}
	\begin{align}
		\hat{H}
		=
		\hat{H}^e
		+
		\hat{H}^c
		+
		\hat{H}^{int}
		+
		\hat{H}^{dse}
		\quad,
		\label{eq.pauli_fierz_hamiltonian}
	\end{align}
	in dipole approximation, length gauge representation and {for fixed} nuclei.  The first term is the bare electronic Hamiltonian, ${\hat{H}^e=\hat{T}_e+V_{ee}+V_{en}+V_{nn}}$, which accounts for the electronic kinetic energy operator, $\hat{T}_e$, Coulomb interactions between electrons ($V_{ee}$, in case of more than one electron), the electron-nuclear interaction ($V_{en}$), and, in case of molecules, the internuclear repulsion ($V_{nn}$). 
\\
	
	The second term is an effective single-mode cavity Hamiltonian
	\begin{align}
		\hat{H}^c
		&=
		\hbar\omega_c
		\left(
		\hat{a}^\dagger_c
		\hat{a}_c
		+
		\dfrac{1}{2}
		\right)
		\quad,
	\end{align}
	with cavity frequency, $\omega_c$, besides photon creation and annihilation operators, $\hat{a}^\dagger_c$ and $\hat{a}_c$, respectively. 
	The quantum light-matter interaction term is given by
	\begin{align}
		\hat{H}^{int}
		&=
		g_c\,
		\left(
		\underline{e}_\lambda
		\cdot
		\underline{\hat{\mu}}
		\right)
		\left(
		\hat{a}^\dagger_c
		+
		\hat{a}_c
		\right)
		\quad,
	\end{align}
	with effective light-matter interaction constant, $g_c$, and cavity polarization vector, $\underline{e}_\lambda$, with polarization direction, $\lambda$. 
	Further, $\underline{\hat{\mu}}$ is the 
	atomic / molecular dipole operator, composed of electronic ($e$) and 
	nuclear parts ($n$), {\em i.e.}, ${\underline{\hat{\mu}}=\underline{\hat{\mu}}_e+\underline{\hat{\mu}}_n}$. The interaction parameter, 
	$g_c$ (formally in units of an electric field strength), can be explicitly written as
	\begin{align}
		g_c
		&=
		\sqrt{\dfrac{\hbar\omega_c}{2\varepsilon\,V_c}}
		\quad,
	\end{align}
	where $\varepsilon$ is the permittivity of 
	the material within the cavity, and $V_c$ is an effective cavity volume. In this work, we will treat $g_c$ as a free 
	parameter to investigate different coupling regimes. Note, we will 
		use $g_c$  values which  
		are typically too large for a single atom / molecule, 
		so the couplings should be considered 
		to rather describe the effective interaction of a 
		cavity mode with an ensemble of 
		emitters. The fourth and last term in Eq.(\ref{eq.pauli_fierz_hamiltonian}) is quadratic in $g_c$ and known as the dipole self-energy (DSE)
	\begin{align}
		\hat{H}^{dse}
		&=
		\dfrac{g^2_c}{\hbar\omega_c}
		\left(
		\underline{e}_\lambda
		\cdot
		\underline{\hat{\mu}}
		\right)^2
		\quad .
	\end{align}
	%
	\subsection{Time-dependent Schr\"odinger equation and laser-pulse excitation}
	The time-evolution of the light-matter hybrid system as initiated by a classical laser-pulse excitation of the 
	atom / molecule is studied {\em via} the time-dependent Schr\"odinger equation (TDSE)
	\begin{align}
		{i}\hbar
		\dfrac{\partial  {\Psi(t)}}{\partial t}
		&=
		\left(
		\hat{H}
		-
		\underline{\hat{\mu}}
		\cdot
		\underline{F}(t)
		\right)
		{\Psi(t)}
		\quad,
		\label{eq.tdse}
	\end{align}
	with initial state, ${\Psi(t=0)}$, to be specified below. Here, $\hat{H}$ is the polaritonic Pauli-Fierz Hamiltonian in Eq.\eqref{eq.pauli_fierz_hamiltonian} and the second term refers to a classical laser field with electrical field amplitude, $\underline{F}(t)$, coupled to the molecular subsystem {\em{via}} the corresponding dipole operator, $\underline{\hat{\mu}}$. 
	We note that the classical laser field does not couple to the cavity mode.
\\
	
	In all applications below, the driving classical field is modelled as a pulse of the form
	\begin{align}
		\underline{F}(t)
		=
		F_0
		\ \underline{P}
		\ \cos(\omega_0 (t-t_p))
		\ \cos^2\left(\frac{\pi}{2\sigma_p}(t-t_p)\right)
		\quad,
	\end{align}
	with peak field intensity, $F_0$, field polarization vector, $\underline{P}$, and peak time with maximal amplitude at half the pulse length, $t_p=\sigma_p$, respectively. The pulse starts at $t=0$ and ends at $t_f=2\sigma_p$.
In all cases below we shall use pulses linearly 
 polarized
	along the  
	electron coordinate in the 1D atomic model, or 
	the molecular axis in case of the molecular model.
\\

	We solve Eq.\eqref{eq.tdse} either for a one-dimensional H atom model or a H$_2$ molecule with fixed orientation and bond length {\em via} grid techniques (for the atom) or the QED-TD-CIS method employing a polaritonic basis (\textit{cf.} Sec.\ref{subsec.polariton_dynamics}).
	\subsection{Calculation of HHG signals and other properties} 
	\label{props}
	HHG spectra are obtained {\em{via}} the Fourier transform of the dipole acceleration function of the $z$-component of the dipole operator (which is also the classical-field polarization direction), 
	computed as $\mu_z(t) = \langle\Psi(t)|\hat{\mu}_z|\Psi(t)\rangle$ from the time-dependent polaritonic wavefunction, 
	$\Psi(t)$, according to
	\begin{align}
		I(\omega) 
		\propto
		\left
		\vert
		\int_0^{t_f}
		w(t) 
		e^{-i\omega t}
		{\left(\frac{\partial^2\mu_z(t)}{\partial t^2}\right)}
		dt 
		\right\vert^2
		\quad .
		\label{hhgsig}
	\end{align}
	In practice, we  employ a Hann window in the integrand, $w(t)=\sin^2(\pi t/ t_f)$, to reduce noise in the simulated spectra.
	For the calculation of HHG spectra, the wave function is not renormalized during the propagation, taking into account ionization losses (see below).
	\\
	
	To ``measure'' the cutoff of computed 
 HHG spectra, a least-squares fit was applied after taking a smooth average of the spectrum, ${\bar{I}(\omega) = \int_{\omega-\Delta \omega}^{\omega_+\Delta\omega}I(\omega')d\omega'}$ (with an averaging width $2\Delta\omega\approx4\omega_0$). The target function to be fitted, describes a constant plateau, linear descend and a constant cutoff as:
	\begin{align}
		\ln \bar{I}(\omega) \approx \left\{ \begin{array}{r l}
		A & \textrm{for } \omega < \omega_a \\
		A-(A-B)\frac{\omega-\omega_a}{\omega_b-\omega_a} & \textrm{for } \omega_a < \omega < \omega_b \\
		B & \textrm{else} \quad .
	\end{array} \right.~, 
\end{align}
	The cutoff can therefore be described 
	by three quantities, the end of the plateau, $\omega_a$, the start of the noise threshold, $\omega_b$, and the middle of the cutoff, ${\omega_\mathrm{cut}=\frac{1}{2}(\omega_a+\omega_b)}$.
	\\
	
	Another property
	of interest is the cavity 
	photon number expectation value, which takes in length gauge representation 
	the form\cite{Fischer2023}
	\begin{align}
		\braket{
			\hat{n}_c
		}
		(t)
		=
		\dfrac{1}{\hbar\omega_c}
			\braket{
				\Psi(t)
				\vert
				\left(
				\hat{H}^c
				+
				\hat{H}^{int}
				+
				\hat{H}^{dse}
				\right)
				\vert
				\Psi(t)
		}
		-
		\dfrac{1}{2}
		\quad.
		\label{nc}
	\end{align}
	
	\section{Models and Numerical Methods}
	\label{sec3}
	\subsection{One-dimensional H atom in a cavity}
	\subsubsection{The model}
	As a starting point for disentangling the details of the HHG process in presence of the cavity field, we employ a simple two-dimensional model Hamiltonian. Specifically, the matter subsystem is composed of a one-dimensional electron coupled to a single fixed nucleus with effective electronic Hamiltonian  
	\begin{align}
		\hat{H}^e
		&=
		-
		\dfrac{\hbar^2}{2}
		\dfrac{\partial^2}{\partial z^2}
		- 
		\frac{Z}{\vert z-R\vert +\eta}
		\quad,
	\end{align}
	with electronic coordinate, $z$,  
	nuclear charge, $Z$, and screening parameter, $\eta$, which regularizes the Coulomb cusp of the interaction potential. Additionally, we fix the nucleus at the origin, {\em i.e.}, the nuclear position is chosen as $R=0$. In presence of a classical, driving laser field coupled to the matter subsystem, $\hat{H}^e$ is extended as $\hat{H}^e - \hat{\mu}_z F_z(t)$, where $\hat{\mu}_z=-z$ is the 
 (electronic) dipole moment along $z$.
	\\
	
	Here, we choose a coordinate representation of the harmonic cavity mode by introducing a ``cavity displacement coordinate'', $x_c$, and its conjugated momentum 
	operator, $\hat{p}_c$\cite{Fischer2021} 
	\begin{align}
		\label{xc}
		x_c
		&=
		\sqrt{\dfrac{\hbar}{2\omega_c}}
		\left(
		\hat{a}^\dagger_c
		+
		\hat{a}_c
		\right)
		\quad,
		\vspace{0.2cm}
		\\
		\hat{p}_c = 
		-
		{i}\hbar
		\dfrac{\partial}{\partial x_c}
		&=
		-
		{i}
		\sqrt{\dfrac{2\omega_c}{\hbar}}
		\left(
		\hat{a}^\dagger_c
		-
		\hat{a}_c
		\right)
		\quad.
	\end{align}
	The cavity Hamiltonian and the light-matter interaction term then take the form
	\begin{align}
		\hat{H}^c
		&=
		-
		\dfrac{\hbar^2}{2}
		\dfrac{\partial^2}{\partial x^2_c}
		+
		\dfrac{\omega^2_c}{2}
		x^2_c
		\quad,
		\vspace{0.2cm}
		\\
		\hat{H}^{int}
		&= 
		-
		\sqrt{\dfrac{2\omega_c}{\hbar}}
		g_c\,
		z
		x_c
		\quad ,
	\end{align}
{\em i.e.}, the cavity mode is chosen to be polarized along $z$.
	Finally, the DSE Hamiltonian reads in this model,
	\begin{equation}
		\hat{H}^{dse} 
		=
		\frac{g_c^2}{\hbar \omega_c}
		z^2
		\quad ,
		\label{dse}
	\end{equation}
	which resembles a light-matter coupling dependent harmonic confining potential along the electronic coordinate
	\subsubsection{Numerical realization}\label{sec:1e-numreal}
	In order to solve the classically-driven polaritonic TDSE, we employ a grid representation. First, the electronic potential in 
		$\hat{H}^e$ was constructed with a regularization constant $\eta=0.9871$ $a_0$ and an effective nuclear charge $Z=1$. We then diagonalize the field-free 
		electronic Hamiltonian {\em via} a Fourier Grid technique\cite{Kosloff1988, Marston1989}, using 
 an equidistant 
        grid
 over 512 points between $z \in [-100,+100]$ $a_0$.
 This 
 provides field-free
 electronic eigenvalues, the lowest three ones given as
		$E_0=-0.500008$, $E_1=-0.1815345$, and $E_2=-0.112529$ (all in Hartree, $E_h$), compared to the analytic eigenvalues of a 3D H atom 
		of $E_0=-0.5$, $E_1=-0.25$, and $E_2=-0.11111$ $E_h$.
		These 1D hydrogen energies were the result of optimizing the regularized potential 
 to achieve the same ionization potential, as well as similar behavior of other electronic energies compared with the 3D system.
	\\
	
	For solving the polaritonic TDSE, 
 a time-dependent version of the Fourier grid Hamiltonian technique is used\cite{sathyamurthy2021time} with the same $z$-grid as used for stationary calculations and 
an equidistant grid 
 chosen also for the cavity coordinate, $x_c$. 
 The latter consists  
	of $N_p$ points between $\pm x_{max}$, 
	where $N_p$ and $x_{max}$ depend on the chosen 
	cavity frequency, $\omega_c$ (see below). 
	Further, in order to avoid reflections of the polaritonic 
	wave packet from grid boundaries, 
	spatial complex absorbing potentials 
	(CAPs) $\Gamma$ were added to the total
	Hamiltonian for electron and cavity coordinates:
	\begin{align}
		-{i}
		\Gamma(z,x_c)
		&=
		- {i}
		\left(
		\Gamma(z)
		+
		\Gamma(x_c)
		\right)
		\quad.
	\end{align}
	The electronic and photonic CAPs, $\Gamma(z)$ 
 and $\Gamma(x_c)$, were chosen 
 as specified in Appendix A and Tab.\ref{tab1} there.
 For different 
 cavity frequencies, $\omega_c$, different 
 CAPs $\Gamma(x_c)$ and also cavity grid sizes (governed by $x_{max}$) had to be chosen, which is 
 also described in Appendix A.
	\\

Two-dimensional polaritonic wavepackets $\Psi(z,x_c;t)$ 
 were propagated with a 4$^\mathrm{th}$-order Runge-Kutta algorithm, representing the electronic wave function on the chosen ($z$, $x_c$) grid and Fast Fourier Transform (FFT) techniques are used to apply the kinetic terms efficiently.\cite{sathyamurthy2021time}
	The time step was chosen as $\Delta t = 0.001~\hbar/E_h$ (0.024~as).
	As initial state, $\Psi(t=0)$, we take the polaritonic ground state of the laser-field free polaritonic Pauli-Fierz Hamiltonian, which was obtained {\em via} imaginary time propagation. 
	\subsection{H$_2$ in a Cavity and QED-TD-CIS}
	\label{subsec.polariton_dynamics}
	\subsubsection{TD-CIS}
	The electron dynamics of a molecule driven by a classical laser field can be described by means of the TD-CI method, here used in its configuration interaction singles form (TD-CIS).  In TD-CIS, the time-dependent electronic wave packet, ${\Psi(t)}$, is expanded in a basis of CIS states, ${\Phi_i}$, with time-dependent coefficients, $C_i(t)$, as \cite{Klamroth2003,Krause2005,Santra2006}
	\begin{align}
		{\Psi(t)}
		&= 
		\sum_i 
		C_i(t)
		{\Phi_i} 
		\quad,
\vspace{0.2cm}
\\
		{\Phi_i} 
		&= 
		D_{0,i} {\Psi_0} +
		\sum_a^{occ} 
		\sum_r^{virt} 
		D_{a,i}^r 
		{\Psi_a^r}
		\quad.
		\label{eq.td_ci}
	\end{align}
	The CIS states ${\Phi_i}$ in Eq.(\ref{eq.td_ci}) are represented in a basis spanned by the Hartree-Fock reference state, ${\Psi_0}$, and singly-excited Slater determinants, ${\Psi_a^r}$, respectively. 
	The latter are obtained from single-excitations 
	of ${\Psi_0}$, by excitation 
 of an electron from occupied orbital $a$ to  
 virtual orbital, $r$. Expansion coefficients are $D_{0,i}$ for 
 the reference determinant, and $D_{a,i}^r$ for singly excited determinants, where indices $a$ and $r$ run an occupied (occ) molecular orbital (MO) a in $\Psi_0$ to a virtual (virt) MO r.
 Expansion coefficients are obtained from solving the field-free, CIS eigenvalue problem.
	\\
	
	A time-dependent classical laser field, $\underline{F}(t)$, leads to interstate couplings mediated {\em via} dipole operator matrix elements, $\underline{\mu}_{ij}$, such that
	\begin{align}
		H^e_{ij}(t) 
		= 
		\left(
		E_i 
		-
		\frac{{i}}{2}
		\Gamma_i 
		\right) 
		\delta_{ij}
		-
		\underline{\mu}_{ij} 
		\cdot
		\underline{F}(t)
		\quad 
		\label{heij}
	\end{align}
	with CIS eigenenergy $E_i$ corresponding to CIS state $\Phi_i$, respectively.
	In order account for possible, 
 partial ionization of the molecule due to the laser field, CIS energy of the ith electronic state, $E_i$, is here augmented by an imaginary term with state-specific ionization rate, $\Gamma_i$, which is chosen according to a heuristic ionization model as\cite{Klinkusch2009}
	\begin{align}
		\Gamma_i 
		= 
		\begin{cases}
			0 &,\quad\text{if}~ E_{i} < I_p 
			\vspace{0.2cm}
			\\
			\displaystyle\sum_a^{occ} \displaystyle\sum_r^{virt}  |D_{a,i}^r|^2	\dfrac{ \sqrt{\varepsilon_r}}{d} &,\quad \text{else} \quad .
		\end{cases}
		\quad.
	\end{align}
	Here, $I_p=-\varepsilon_{\mathrm{HOMO}}$ is the ionization potential as obtained from Koopmans' theorem; $\varepsilon_r$ is the 
	energy of the $r^\mathrm{th}$ virtual MO, 
	which relates to the excess energy of an escaping electron,
	and $d$ is an ``escape length'' parameter, beyond which the electron is considered ``free''.
	\subsubsection{QED-TD-CI}
	An extension of TD-CI to the ESC regime has been recently introduced in terms of the quantum electrodynamical time-dependent configuration interaction method (QED-TD-CI)\cite{qedtdci}. Here, the time-dependent polaritonic wave packet is expanded in a basis of polaritonic states, ${\Phi_{p}^\mathrm{QED}}$, as
	\begin{align}
		{\Psi^\mathrm{QED}(t)}
		&=
		\sum_p C^\mathrm{QED}_p(t)
		\ {\Phi_{p}^\mathrm{QED}}~,
	\end{align}
	which are obtained by diagonalizing the polaritonic Pauli-Fierz Hamiltonian (\ref{eq.pauli_fierz_hamiltonian}). 
	For this purpose, one chooses a basis of 
	zero-order product states, ${\Phi_i} {\psi_n}$, composed of electronic states, ${\Phi_{i}}$, diagonalizing the CIS Hamiltonian,
  and photonic states, ${\psi_n}$, which are eigenstates of $\hat{H}_c$. 
	The polaritonic states, are expanded in $N_\mathrm{CI}$ CIS states and $N_{p}$ photon states as
	\begin{align}
		{\Phi_{p}^\mathrm{QED}}
		&=
		\sum_{i=0}^{N_\mathrm{CI}-1} 
		\sum_{n=0}^{N_p-1} 
		D^\mathrm{QED}_{p,in}
		\ {\Phi_i}
        {\psi_n}
		\quad .
		\label{eq.qed_td_ci}
	\end{align}
	In this zero-order basis, the Hamiltonian matrix takes the form
	\begin{align}
		H_{in, jm}
		&= 
		H^e_{in,jm}
		+
		H^c_{in,jm}
		+
		H^{int}_{in,jm}
		+
		H^{dse}_{in,jm}
		\quad,
	\end{align}
	where the individual terms are explicitly given as
	\begin{align}
		H^e_{in,jm}
		&=
		E_{i}
		\ \delta_{ij}
		\ \delta_{nm} 
		\quad,
		\vspace{0.2cm}
		\\
		H^c_{in,jm}
		&=
		n \ \hbar \omega_c
		\ \delta_{ij}
		\ \delta_{nm}
		\quad,
		\vspace{0.2cm}
		\\
		H^{int}_{in,jm}
		&=
		g_c 
		(\underline{e}_c\cdot\underline{\mu}_{ij})
		(
		\sqrt{n}\,
		\delta_{m,n-1} 
		+
		\sqrt{m}\,
		\delta_{m,n+1}) 
		:=
		c_{ij}
		(
		\sqrt{n}\,
		\delta_{m,n-1} 
		+
		\sqrt{m}\,
		\delta_{m,n+1}) 
		\quad,
		\vspace{0.2cm}
		\\
		H^{dse}_{in,jm}
		&=
		\frac{g_c^2}{\hbar \omega_c} 
		\sum_k
		\left( 
		(\underline{e}_c \cdot \underline{{\mu}})_{ik}
		\cdot
		(\underline{e}_c \cdot \underline{{\mu}})_{kj} \right)\,
		\delta_{nm}
		:=
		d_{ij}
		\ \delta_{nm}
		\quad.
	\end{align}
	On the diagonal, the CI eigenenergies are augmented by the cavity energy, $n\hbar\omega_c$, where $n$ is the number of cavity photons in the non-interacting limit $(g_c=0)$. 
	Further, light-matter interaction is mediated by the off-diagonal coupling terms, $c_{ij}$, and the block-diagonal terms, $d_{ij}$, related to the DSE term.
	\\

		The laser-driven system is then propagated in the basis of polariton eigenstates with Hamiltonian matrix elements
		\begin{align}
			H^e_{pq}
			=
			E_{p}
			\ \delta_{pq}
			- 
			\underline{\mu}_{pq} \underline{F}(t)
			\quad ,
		\end{align}
		with polaritonic eigenenergy $E_p$ of state $\Phi^{QED}_p$ and dipole moments obtained from the corresponding transformation of the molecular dipole matrix as
		\begin{align}
		\underline{\mu}_{pq}
		&=
		\sum_{ij,nm}
	    D^\mathrm{QED}_{p,in}
		D^\mathrm{QED}_{q,jm}
		\underline{\mu}_{ij}
		\quad.
		\end{align}
	In order to be able to describe ionization from a polariton state, one has to extend the heuristic ionization model 
	for TD-CI, to QED-TD-CI. The most straightforward method, which is 
		adopted here, is 
	to use a weighted sum of the molecular ionization rates, $\Gamma_i$, under the assumption, that the cavity does not directly change the lifetimes of the underlying zero-order states:
	\begin{align}
		\Gamma_{p}^\mathrm{QED} &= 
		\sum_{n i} \vert D_{p,in}^\mathrm{QED}\vert^2 \ \Gamma_i \quad.
	\end{align}
	We note that the question of how molecular ionization should change inside a cavity is still an open discussion in the literature.\cite{cavityionization, Riso2022} Therefore, we assume here a minimal influence and accordingly do not change ionization parameters compared with earlier works.\cite{Klinkusch2009,bedurke2019discriminating,bedurke2021many}
	
	\subsubsection{Numerical realization}
	For the hydrogen molecule, we used
	a fixed bond length of $R=1.4~$a$_0$ ($R=0.741$~\AA) and 
	performed a CIS calculation using the aug-cc-pVTZ basis set and eight Kaufmann shells\cite{Kaufmann1989}, removing  eigenvectors of the orbital overlap matrix with eigenvalues below $10^{-6}$ {\em via} canonical orthogonalization.\cite{SzaboOstlund}
	This results in 113 CIS states 
	and an ionization potential of 0.594
	$E_h$ according to Koopmans' theorem. 
	~$E_h$.
	This basis of CIS states is sufficient to calculate HHG spectra in the non-cavity case in line with previous studies.\cite{bedurke2019discriminating,saalfrank2020molecular,bedurke2021many}
	\\

	Analogous calculations were done for H$_2$ in a cavity using QED-CIS with twenty photons ($N_p=20$) resulting in $20 \times 113=2260$ polariton states. 
	In Tab.\ref{tab2} (\textit{cf.} Appendix~A), we compare energies of the ten lowest lying adiabatic states obtained from CIS and QED-CIS approaches, 
  for a single cavity mode polarized along the 
 molecular axis, with a cavity-molecule coupling constant 
 of $g_c=0.01\,{E_h}/ea_0$. 
 We consider two scenarios with different cavity frequencies, namely 
  $\omega_c=0.057\,E_h$ (which is resonant 
 to an 800 nm laser pulse later used for HHG),
  and $\omega_c=0.467\,E_h$ (corresponding to the first electronic transition 
 energy, $E_1-E_0$ of the field- and cavity-free H$_2$ molecule). Additionally, we show in Fig.\ref{fig:polaritons} the polaritonic character of the QED-CIS eigenstates by the population of its contributing zero-order states in Appendix A. 
 From Tab.\ref{tab2} and Fig.\ref{fig:polaritons} one notes that  
 in the low-frequency scenario, the lowest states (up to the eighth excited state) exhibit mainly cavity-mode excitations, whereas in case of the electronic resonance strong mixing 
 between electron and photon modes occurs and the first and third excited states, for example, form lower and upper polaritons.
	\\

	For (QED)-TD-CIS,
	the coefficient vector was propagated with the first-order split-operator method with time steps ${\Delta t=0.02~\hbar/E_h}$ (0.5~as) over the duration of the laser pulse (${1103.16~\hbar/E_h}$, 26.7~fs), using the QED-TD-CIS  or TD-CIS-Hamiltonian (for the non-cavity case), $\underline{\underline{H}}(t)$, 
	\begin{align}
		\underline{C}^{\mathrm{(QED)}}(t+\Delta t) 
		&= 
		e^{-{i}\underline{\underline{H}}(t)\Delta t/\hbar} 
		\underline{C}^{\mathrm{(QED)}}(t)
		\quad.
		\label{tdci-tdse}
	\end{align}
	
	\section{Results and discussion}
	\label{sec.results}
	\subsection{Hydrogen-atom like one-electron model}
	In a first step, we discuss HHG spectra for the hydrogen-atom like model under ESC. 
	To compensate for the influence of ionization on observables and clearly separate effects of strong coupling from ionization effects, we divide all expectation values (except the dipole acceleration for the calculation of HHG spectra) by the norm, 
	$N(t)=\braket{\Psi(t) \vert \Psi(t)}$.
	We employ a 10-cycle laser pulse with carrier frequency of $\omega_0=0.05$ (in atomic units, $E_h/\hbar$, corresponding to 911 nm) and peak intensity of $0.09~\frac{E_h}{ea_0}$. The frequency was chosen, close to an 800~nm pulse, but rounded for practical convenience.
	\\
	
\begin{figure}[ht!]
		\centering
		\renewcommand{\baselinestretch}{1.}
		\hfill a)\hfill~\\
		\hspace*{1.cm}\includegraphics[width=0.7\linewidth]{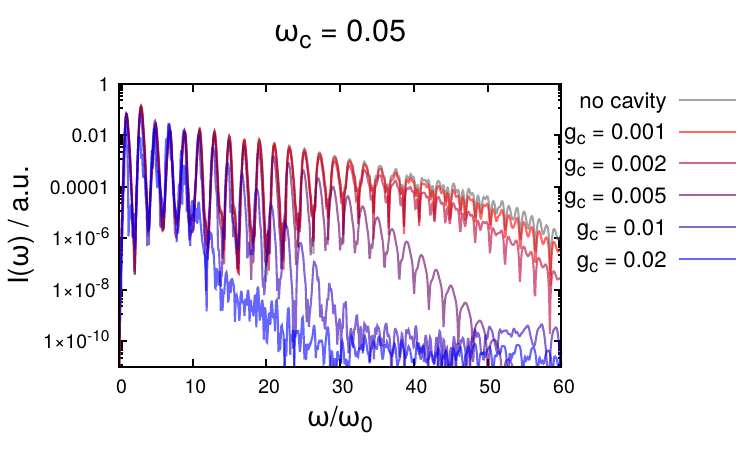}\\
		\hfill b)\hfill~\\
		\hspace*{1.cm}\includegraphics[width=0.7\linewidth]{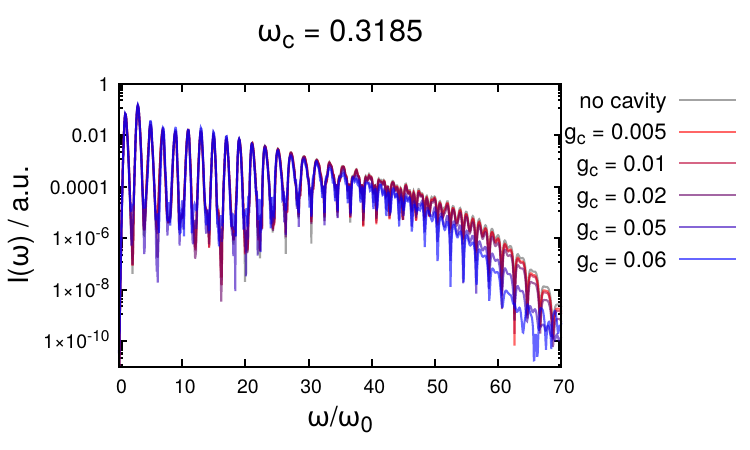}\\
\caption{HHG spectra of a hydrogen-like atom in a cavity during an $\omega_0=0.05~E_h/\hbar$, 10-cycle pulse with $0.09~\frac{E_h}{ea_0}$ peak amplitude, obtained {\em via} grid solution.
			a) HHG spectrum with cavity frequency $\omega_c=0.05~E_h/\hbar$.
			b) HHG spectrum with $\omega_c=0.3185~E_h/\hbar$.
		}\label{h-atom-hhg}
	\end{figure}

	In Figs.\ref{h-atom-hhg}a) and b), the effect of the cavity on HHG spectra is depicted for increasing light-matter coupling strength, $g_c$ (in atomic units, $E_h/(e a_0)$),  a) for two different cavity frequencies, $\omega_c=0.05$ (resonant to the laser frequency) and b) $\omega_c=0.3185$ (resonant to the first, field-and cavity free electronic transition, $\hbar \omega_c=E_1-E_0$, see above Sec.\ref{sec:1e-numreal}), respectively. 
	With increasing $g_c$, we observe a decrease in both intensity of higher harmonics and the cutoff. 
	For very large values of $g_c$, we additionally observe modifications of lower harmonics.
	These effects are most pronounced for the case when the cavity frequency is resonant with the classical-laser frequency ($\omega_c=\omega_0=0.05$, 	see Fig.\ref{h-atom-hhg}a)).
	There, also the 5$^\mathrm{th}$-9$^\mathrm{th}$ harmonic of the case $g_c=0.01$ exceed the intensity of the non-cavity spectrum. 
	The effect becomes significantly weaker for the higher cavity frequency (Fig.\ref{h-atom-hhg}b)). 
	Closer inspection shows 
 that indeed, we see
  only an effect when the cavity frequency is not resonant with any of the transitions between the energy levels of the field-free 1D H atom (see above).
 In particular in Fig.\ref{h-atom-hhg}b), we also do not observe a change in the position of the harmonics as described in the literature.\cite{Aklilu2024}
	This might be attributed to the fact that we employ ionization models, which reduce artifical reflections of the polaritonic wave packet on the grid boundaries.
	\\
	
\begin{figure}[ht!]
		\centering
		\renewcommand{\baselinestretch}{1.}
		\hfill a)\hfill~\\
		\hspace*{1.8cm}\includegraphics[width=0.7\linewidth]{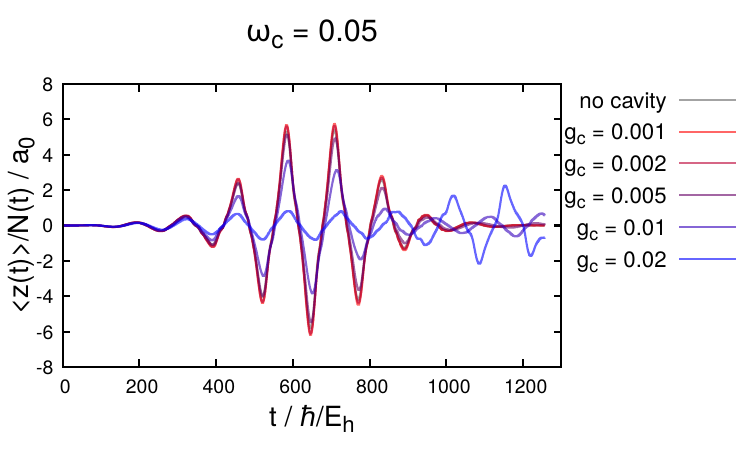}\\
		\hfill b)\hfill~\\
		\hspace{.5cm}
		\hspace*{1.3cm}\includegraphics[width=0.72\linewidth]{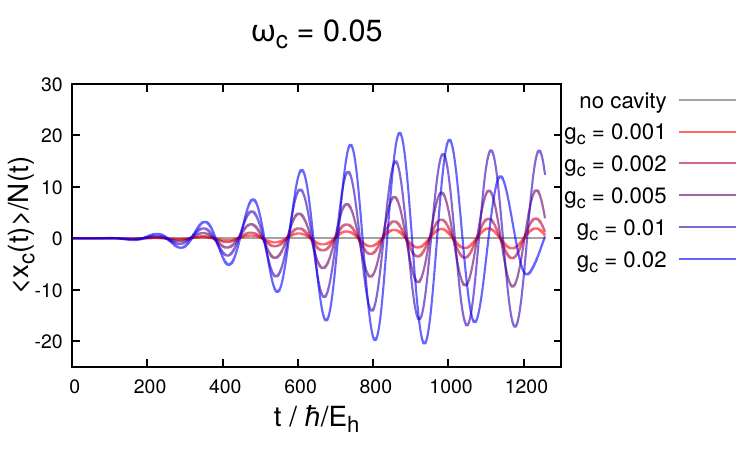}\\
		\caption{Grid solution of an hydrogen-like atom in a cavity during an $\omega_0=0.05~E_h/\hbar$, 10-cycle pulse with $0.09~\frac{E_h}{ea_0}$ peak amplitude. 
			a) Normalized expectation value of the electronic coordinate.
			b) Normalized expectation value of the cavity coordinate.
		}\label{h-atom-hhg-exp}
	\end{figure}

    In order to gain insight into the lowered cutoff, we study the motion of the classical-laser driven electron in the cavity {\em via} the time-dependent position expectation value $\langle z \rangle(t)$ as shown in Fig.\ref{h-atom-hhg-exp}a). We observe that $\langle z \rangle(t)$ decreases with increasing coupling strength, $g_c$, at a given 
		$\omega_c$. This effect is traced back to the DSE term in Eq.(\ref{dse}), which induces a bound harmonic potential along the electronic coordinate restricting electron motion to smaller amplitudes. As a consequence, HHG intensities are lowered, particularly for the higher harmonics, {\em i.e.}, the cutoff shifts to lower energies, as $\langle z \rangle(t)\propto\mu_z(t)$.
	\\
	
\begin{figure}[ht!]
		\centering
		\renewcommand{\baselinestretch}{1.}
		\hfill a)\hfill~\\
		\hspace{.1cm}
		\hspace*{2cm}\includegraphics[width=0.72\linewidth]{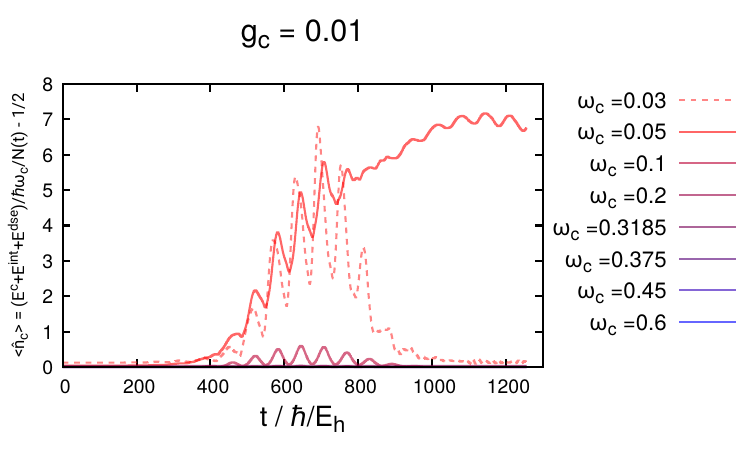} \\
		
		\hspace*{1cm}\includegraphics[width=0.66\linewidth]{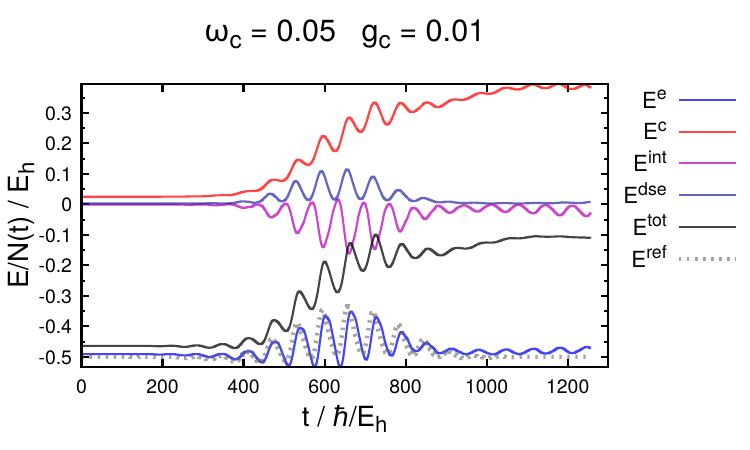} \\
		\caption{Grid solution of an hydrogen-like atom in a cavity during an $\omega_0=0.05~E_h/\hbar$, 10-cycle pulse with $0.09~\frac{E_h}{ea_0}$ peak amplitude. 
			a) Average cavity quantum number dependent on cavity frequency.
			b) Energy distributed between electronic and cavity coordinate for $\omega_c=0.05~E_h/\hbar$ and $g_c=0.07$ and reference electronic energy without cavity.
		}\label{h-atom-hhg-ph}
	\end{figure}

	Additionally, the light-matter hybrid system is continually excited in the cavity coordinate, as observable from the time-dependent expectation value $\braket{x_c}(t)$ shown in Fig.\ref{h-atom-hhg-exp}b). 
There, we observe oscillations,	which increase with increasing coupling strength, $g_c$. 
	The excitation of the cavity mode can be rationalized by inspecting the photon number expectation value, $\langle {\hat{n}}_c \rangle(t)$ ({\em cf.} Eq.(\ref{nc})), as shown in Fig.\ref{h-atom-hhg-ph}a), which increases significantly for a given coupling strength, $g_c$,  when the cavity frequency is equal to the classical-laser frequency, $\omega_0=0.05$.
	Interestingly, there is also some temporary, 
 non-zero photon number present for different frequencies, $\omega_c$, above and below the laser frequency, which do not survive until the end of the pulse.
Seemingly, we excite several photon states, during the laser pulse, without without the classical laser field being directly coupled to the cavity states.
\\

	A connection between both observations, the restriction of electronic motion (seen as the lower-amplitude motion in $z$) and the excitation of the cavity, is obtained by an energy decomposition of the light-matter hybrid system 
 as provided in 
	Fig.\ref{h-atom-hhg-ph}b), for the case
 $\omega_c=0.01$ and $g_c=0.01$ (atomic units). There, the electronic energy, $E^e$ (blue), is shifted to slightly higher energies in the cavity, before the interaction with a laser, if compared to the non-cavity case, $E^{ref}$ (dotted curve).
	During the pulse, {both energies are oscillating towards higher values, 
 with $E^e$ being slightly lower than the reference, non-cavity energy, 
 around the pulse maximum at $\sim 600$ $\hbar/E_h$.
	{This lowering in energy goes hand-in-hand with an increase of the 
 DSE, $E^{dse}$, which oscillates, starting from zero, with positive values during the pulse, in line with the largest amplitude motion of the electron.
	The coupling term, $E^{int}$, shows negative values with an increase during the laser pulse, complementing the DSE term, and remaining oscillation at the end of the propagation.
	The interaction term facilitates a continuous exchange of energy and oscillation between electron ($E^e$) and the cavity.
	This results in a steadily increasing cavity energy, $E^c$:
	The energy of the cavity starts at approximately $\hbar \omega_c/2 = 0.025~E_h$ at $t=0$ and oscillates,
	accumulating energy, and remains oscillating at the end of the pulse, together with $E^e$ and $E^{int}$.
	The total energy, $E^{tot}=E^e+E^{dse}+E^{int}+E^c$, starts elevated by the cavity energy at the beginning and oscillates to higher energies also, mainly driven by the increase in $E^c$.}
	\\

                \begin{figure}[hbt]
                        \centering
                        \renewcommand{\baselinestretch}{1.}
                        \includegraphics[width=0.5\linewidth]{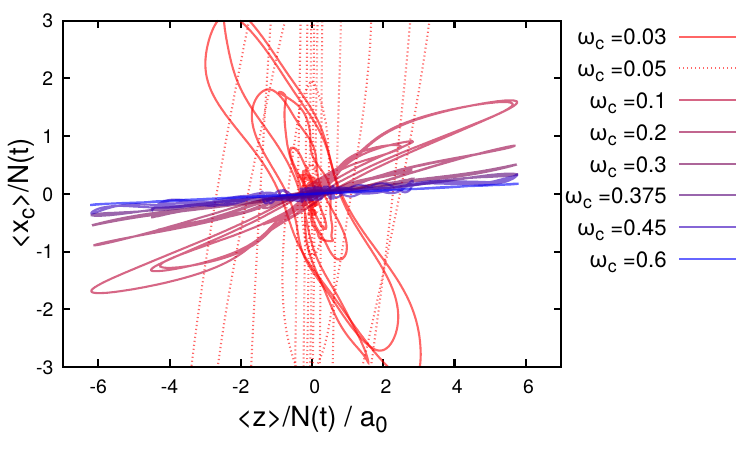}
                        \includegraphics[width=0.48\linewidth]{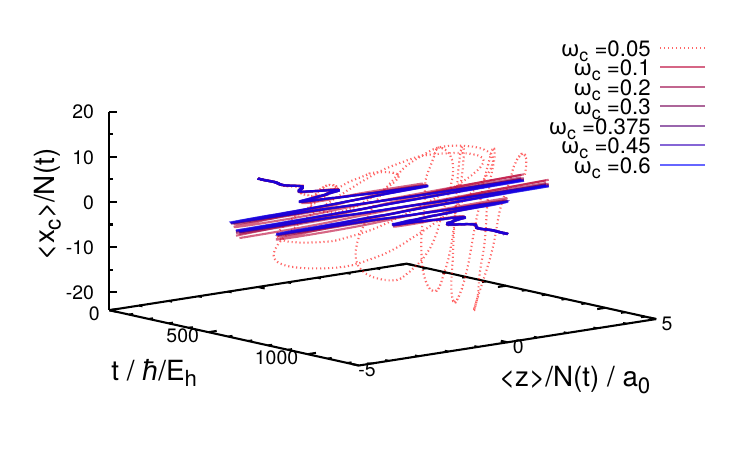}
                        \caption{Normalized displacement coordinate $\langle x_c \rangle/N$ with respect to the normalized electron coordinate $\langle z\rangle/N$ of the 1D H model, showing a phase delay of the properties in a visible angle, which increases when approaching the laser frequency as a resonant case (red, dotted line), which starts vertically and spirals outwards. Left: 2D plot, right: 3D plot.
                        }
                        \label{fig:h-delay}
                \end{figure}

	We would like to again emphasize that the continuous excitation of the cavity is not due to direct laser driving (as the laser does not act on the cavity coordinate in our model), but due to the electron, which first interacts with the electric field and then excites the cavity:
	First the molecular system reacts to the incoming classical 
 field $\underline{F}(t)$ by showing visible oscillations the earliest before the coupling term, $E_{int}$ starts oscillating and leading to the oscillation and excitation of $E_c$ thereafter. More explicitly this becomes apparent from a plot of the time-dependent expectation values for the cavity coordinate $\langle x_c \rangle(t)$ {\em vs.} the electron coordinate $\langle z\rangle(t)$ (see Fig.\ref{fig:h-delay}).
	\\
	
	We observe the response of the cavity coordinate to be delayed with respect to the electron coordinate with a frequency-dependent phase.
	Analogous to coupled oscillations, the delay manifests in an angle between the two coordinates ({\em cf.} Fig.\ref{fig:h-delay}).
	For non-resonant oscillators we would expect no major interaction, {\em i.e.}, an oscillation only around the field acting on the electron along $z$.
	When they become closer in energy, an angle is created, which results at the resonance point in a maximal amplitude along the coupled coordinate, {\em i.e.}, $x_c$.
	For $\omega_c$ below the driving laser frequency, the angle increases further into a negative phase delay, with again less excitation of the cavity.
	In fact according to 
  Fig.\ref{fig:h-delay}, the angle is seen 
 to increase with a lower cavity frequency up to 
 the resonant case, $\omega_c=\omega_0=0.05$. At resonance (red dotted lines), first the oscillation spirals outwards in line with the resonant excitation, so that second a phase shift of $\pi/2$ occurs. This clearly shows the dependence of the cavity excitation on the electronic system, and also shows the DSE-induced low-amplitude electron motion coming from the resonant angle/phase delay. 
	The system remains excited after the pulse, with no possibility to dissipate the energy stored in the cavity.
	\\
	
	In passing, we note that apart from the electron confinement effect in the cavity, a shift of the cutoff 
		could according to Corkum's model (see above) in principle 
		also be caused by an ESC-altered ionization potential, $I_p$, which we do not take into account in this work.
		In fact it has been shown in Ref.\cite{cavityionization}, by 
		QED-CC and QED-HF methods, 
		that ionization potentials can be slightly reduced (for sodium halide compounds), but such a shift (in the order of a few tens of meV) is 
		probably too small to cause large cavity 
		effects on HHG as manifested in Fig.\ref{h-atom-hhg}a) and b).
	\\

	\subsection{Hydrogen molecule: QED-TD-CIS calculations}
	Next, we discuss the HHG spectrum of the hydrogen molecule, coupled to a cavity mode, where we choose cavity frequencies $\omega_c$,  
		varying between the laser frequency $\omega_0=0.057~E_h/\hbar$ (corresponding to 800~nm), and the excitation energy to the first excited state 
		of the free molecule, $(E_1-E_0)/\hbar= 0.467~E_h/\hbar$, ({\em cf.} Tab.\ref{tab2} in Appendix B).
	If not stated 
	otherwise, we used a coupling strength of $g_c=0.01$. 
	\\

	\begin{figure}[hbt]
		\centering
		\hfill a)\hfill~\\
		\hspace*{.5cm}\includegraphics[width=0.75\linewidth]{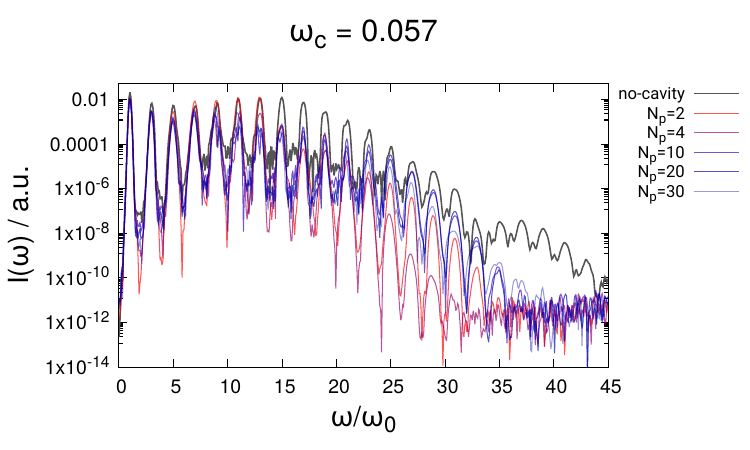}\\
		\hfill b)\hfill~\\
		\includegraphics[width=0.58\linewidth]{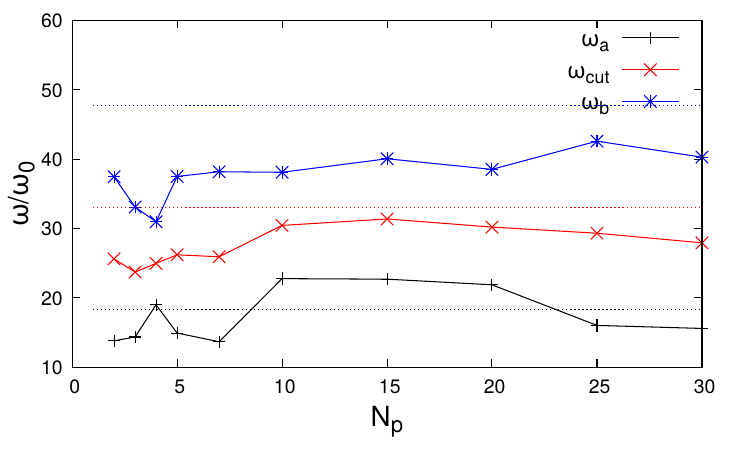}\\
\renewcommand{\baselinestretch}{1.}
		\caption{Hydrogen molecule, H$_2$, in a cavity, 
 treated with the QED-TD-CIS method during an $\omega_0=0.057~E_h/\hbar$, $0.09~\frac{E_h}{ea_0}$ peak amplitude 10-cycle laser pulse. 
			a) HHG spectra with cavity frequency resonant to the classical laser $\omega_c=0.057 E_h/\hbar$ and varying number of photon states, $N_p$. b) Analysis of the cutoff  measures $\omega_a$,
			$\omega_b$, $\omega_\mathrm{cut}$ 
			of the spectra with $\omega_c=0.057 E_h/\hbar$ with non-cavity cutoff parameters shown   
			as horizontal dotted lines.}\label{fig:HHG-tdci-laserres}
	\end{figure}

	In Fig.\ref{fig:HHG-tdci-laserres} and \ref{fig:HHG-tdci-eres}, 
 HHG spectra and their convergence with respect to the number of photon states, $N_p$, used 
 in the expansion of the polaritonic wave packet, $N_p$, are investigated. 
	Compared to 
		the free molecule (``no cavity''), a shift of the 
		HHG plateau cutoff to lower energies / harmonics can be observed, 
		similar to what had been found for the one-dimensional H atom model above. Fig.\ref{fig:HHG-tdci-laserres}a) shows spectra for $\omega_c = 0.057~E_h/\hbar$, resonant to the 800~nm laser: 
	There, it is seen that a large 
	number of photon states (in the order of 20) 
	is needed to converge the HHG spectrum.
	This is also in line with previous observations for the 1D model system: 
	In case of a cavity frequency resonant to the classical laser, higher photon states are populated  which leads to a lower cutoff in the HHG spectrum. 
	We assume this happens through an indirect excitation mechanism {\em via} energy transfer between the electronic and cavity subsystems.
	In Fig.\ref{fig:HHG-tdci-laserres}b), this trend of intensity shifts and convergence is shown with the position of the cutoff plateau region, $\hbar \omega_a$, 
		noise level, $\hbar \omega_b$, and cutoff, 
		$\hbar \omega_\mathrm{cut}$ ({\em cf.} 
		Sec.\ref{props}). 
	These cutoff criteria stabilize only after around 20 photon states, 
	{\em i.e.}, when the maximal cavity energy, $(N_p-1) \hbar \omega_c$, reaches the observed cutoff region in the HHG spectrum, $\hbar \omega_\mathrm{cut}$.
	\\

When setting the cavity frequency to $0.467~E_h/\hbar$, 
{\em i.e.}, resonant to the first excited molecular state, the HHG spectrum converges with only three to four photon states according  to Fig.\ref{fig:HHG-tdci-eres}a). Note that also in this case, 
$(N_p-1) \hbar \omega_c$ is in the region of the HHG cutoff energy.
We conclude that the maximum cavity energy needed to describe the dynamics has to be similar to the maximum energy the molecule 
	emits during HHG, 
	$\hbar\omega_{cut}\approx(N_p-1)\hbar\omega_c$, as this is the energy available 
	after multi-photon excitation during the HHG process.\cite{Corkum2007} 
 We see, with the higher cavity frequency, less changes in the HHG spectrum due to the cavity.
\\

\begin{figure}[H]
	\centering
	\hfill a)\hfill~\\
	\includegraphics[width=0.75\linewidth]{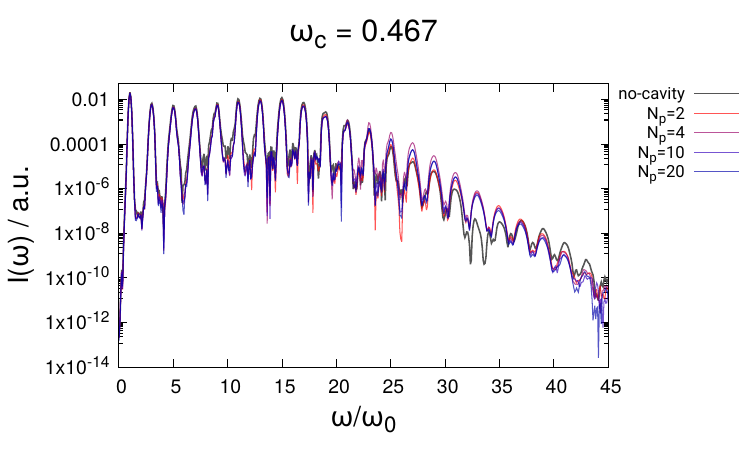}\\
	\hfill b)\hfill~\\
	\includegraphics[width=0.75\linewidth]{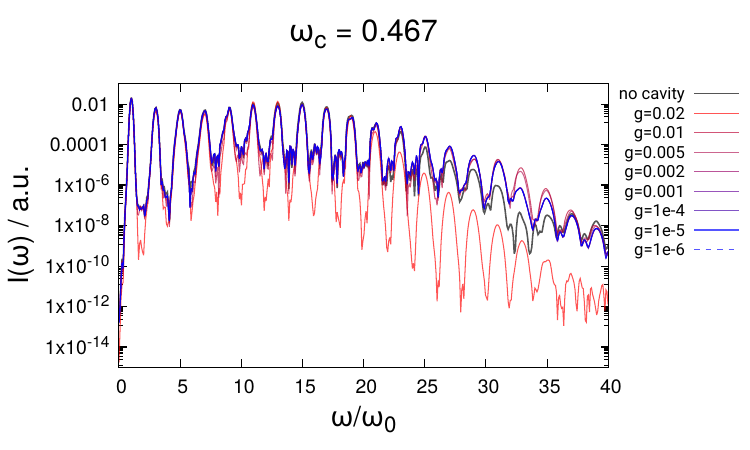}\\
\renewcommand{\baselinestretch}{1.}	
	\caption{
Hydrogen molecule, H$_2$, in a cavity, 
 treated with the QED-TD-CIS method during an
 $\omega_0=0.057~E_h/\hbar$, $0.09~\frac{E_h}{ea_0}$ peak amplitude 10-cycle laser pulse. 
		a) HHG spectra with cavity frequency resonant to the first excited state, $\omega_c=0.467 E_h/\hbar$ and varying number of photon states, 
		$N_p$. b) HHG spectra with cavity frequencies $\omega_c=0.467 E_h/\hbar$ and varying coupling strengths, $g_c$ 
		(for $N_p = 20$).}\label{fig:HHG-tdci-eres}
\end{figure}

Moreover, the HHG spectrum  for $\omega_c= 0.467~E_h/\hbar$ and $g_c=0.01$ in 
Fig.\ref{fig:HHG-tdci-eres}a) is generally 
quite close to the non-cavity spectrum, with the cutoff shift 
	being small.
Main differences are some sharper peaks 
and a ``filled'' irregular shape at the 33th harmonic under ESC. 
	Fig.\ref{fig:HHG-tdci-eres}b), where the HHG spectrum 
	for $\omega_c=0.467~E_h/\hbar$ is shown as a function 
	of the coupling strength, $g_c$, demonstrates that (i) spectra are 
	little affected for $g_c \leq 0.01$, but (ii) 
	show a 
	lowered cutoff for a larger coupling, $g_c=0.02$.
\\

\begin{figure}[H]
	\centering
	\hfill a)\hfill~\\
	\hspace*{.5cm}\includegraphics[width=0.75\linewidth]{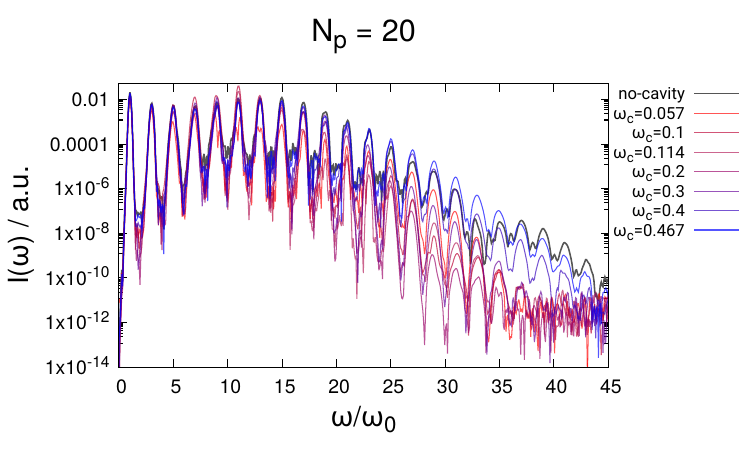} \\
	\hfill b)\hfill~\\
	\includegraphics[width=0.58\linewidth]{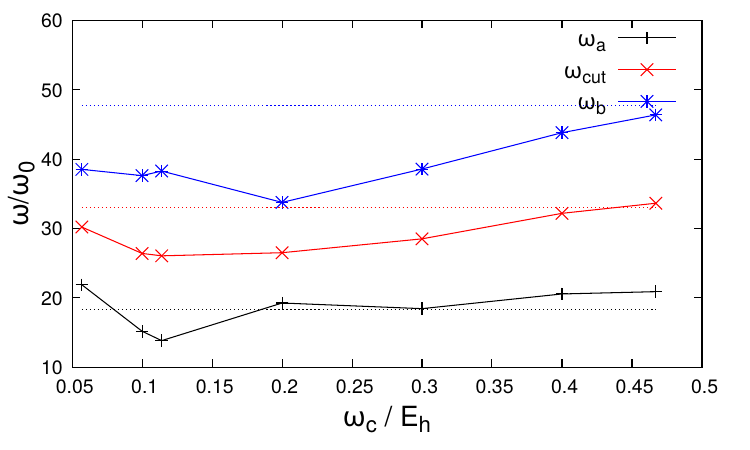}
	\renewcommand{\baselinestretch}{1.}
	\caption{Hydrogen molecule, H$_2$ with the QED-TD-CIS method during a $\omega_0=0.057~E_h/\hbar$, $0.09~\frac{E_h}{ea_0}$ peak amplitude 10-cycle laser pulse. 
		a): HHG spectra for various 
		cavity frequency $\omega_c$ (for $N_p = 20$); b): analysis of the cutoff 
		measures of the spectra with varying cavity frequency with non-cavity cutoffs as horizontal dotted lines.}\label{fig:HHG-tdci-freq}
\end{figure}

Comparing the effect of the different cavity frequencies for 
a fixed (converged) maximal photon number, $(N_p-1)$, more 
systematically, Figs.\ref{fig:HHG-tdci-freq}a) and b) show lower cutoff energies for the lower frequencies with a minimum of the plateau at $\omega_c=\omega_0$.
Interestingly, also some of the harmonics at the plateau are intensified for lower cavity frequencies, {\em e.g.} for 
 $\omega_c=0.1$~({\em cf.} Fig.\ref{fig:HHG-tdci-freq}a)). 
\\

A lowering of the HHG cutoff is in line with the excitation of the cavity 
 and describes an energy transfer during the HHG process, notably 
 at low cavity frequencies.
This energy transfer, reasoned by the indirect excitation pathway, similar to the H atom, is supported by a state population analysis in Appendix C, which reveals non-vanishing polariton state populations 
 for $\omega_c=0.057$ even at the end of the driving laser pulse ({\em cf.} Fig.\ref{t1}). 
 In contrast for 
  $\omega_c=0.467$, this energy transfer is ineffective 
 after the driving laser pulse is off, as can be seen in 
 Fig.\ref{t2}. 

%
\section{Summary and conclusions}
\label{sec.conclusion}
In this work, 
we presented theoretical results using idealized model 
 systems, for HHG spectra of 
atoms / molecules within a cavity. 
We assumed that an atom or molecule
	is driven by an intense classical laser field, giving rise  
	to HHG, while being coupled to the quantized field of an optical cavity.
Specifically, (i) a one-dimensional single-electron model coupled to a single cavity mode 
was considered using grid methods, and (ii) an H$_2$ 
molecule coupled to a cavity mode was
treated by the QED-TD-CIS method. 
\\

The one-electron grid solution shows an indirect excitation of the cavity mode during the HHG process, which leads to lowered intensities of the highest harmonics and 
 a corresponding red-shift of the cutoff frequency. 
 This effect is particular important
 when the laser frequency, $\omega_0$, and cavity frequency, $\omega_c$, differ from each other.
A similar effect is observed for our molecular system, H$_2$, 
with a lowered cutoff, in 
particular when the cavity mode is non-resonant with a molecular 
 excitation energy, {\em e.g.}, when $\omega_c$ was low and close to 
 the laser frequency, $\omega_0$.
 Additionally, the cavity seems to enhance the HHG intensities for these 
 lower cavity frequencies in the plateau region, possibly by redistributing HHG intensity {\em via} excitation and deexcitation of polariton states.
Both effects of the cavity, the cutoff shift and HHG signal intensity variations,
are interesting from the point of view of ``tailoring'' HHG signals, {\em e.g.}, to suppress or amplify selected signals, and this possibility should be further exploited in the future.
{Since the effects observed in this work are strong when the cavity 
 frequency was close to the laser frequency, the question arises to which 
 extent a possible, direct excitation of the cavity by light needs to be included 
 in improved models -- another 
 future research line.}
 Finally, from a methodological 
 viewpoint, 
 the QED-TD-CI method appears as a promising tool, due to its simplicity, to treat ESC 
 also in more complex molecular systems, including ensembles in the future.
\section*{Acknowledgements}
The authors acknowledge funding of this work by the Deutsche Forschungsgemeinschaft (DFG, German Research Foundation) -- CRC/SFB 1636 -- Project ID 510943930 - Project No. A05, ``Understanding and controlling reactivity under vibrational and electronic
strong coupling''. E.W. Fischer acknowledges funding by the Deutsche Forschungsgemeinschaft (DFG, German Research Foundation) through DFG project 536826332.

\section*{Data Availability Statement}
The data that support the findings of this study are available from the corresponding author upon reasonable request.

\section*{Conflict of Interest} 
The authors have no conflicts to disclose.

\section*{Author contributions}

\textbf{Paul A. Albrecht:} Software, Data Curation and Visualization (lead); Methodology (equal); Writing - Original Draft (equal); Writing - review and editing (equal). 
\textbf{Eric W. Fischer:}  Methodology (equal); Writing - Original Draft (equal); Writing - review and editing (equal). 
\textbf{Tillmann Klamroth:} Writing - review and editing (equal); Software (supporting).
\textbf{Peter Saalfrank:} Conzeptualization (lead); Methodology (equal); Writing - review and editing (equal).

\clearpage

\setcounter{equation}{0}
\counterwithin{equation}{subsection}
\renewcommand{\theequation}{\thesubsection.\arabic{equation}}

\section*{Appendices}\label{sec.appendix}
\subsection{One-dimensional H atom in a cavity: Absorbing potentials and grid parameters}
\label{appa}
The function $\Gamma(z)$ which determines the complex absorbing potential (CAP) 
 for the electron coordinate, 
 was chosen as 
\begin{align}
	\Gamma(z) &= \left\{ \begin{array}{r l} 0 &\quad\text{for } z < s  \\ a (z-s)^2 &\quad\text{else} \end{array} \right.~.
\end{align}
A start value of $s=0.67~a_0$ was chosen, with an incline of $a=1.0135\times10^{-4} ~\frac{\text{E}_\text{h}^2}{a_0^2}$. \\

For the 
 cavity  coordinate, $x_c$, we employ a linear CAP function:
\begin{align}
	\Gamma(x_c) &= \left\{ \begin{array}{r l} 0 &~\text{for } x_c < W_s  \\ a_W (x_c-W_s) &~\text{else} \end{array} \right.~.
\end{align}
For the latter, in order to avoid artificial wave packet reflections, 
 parameters and grid extensions ($x_{max}$) had to be chosen to depend 
 on the cavity frequency, $\omega_c$, as detailed in \autoref{tab1}.
\begin{table}[H]
	\centering
	\renewcommand{\baselinestretch}{1.} 
	\caption{Grid and CAP parameters for the cavity coordinate, $x_c$, in the one-electron grid solution (all in atomic units).}
	\begin{tabular}{c | c | c | c | c}
		$\omega_c$ & $N_p$ & $x_{max}$ & $W_s$ & $a_W$ \\
		\hline
		0.03 & 256 & 65 & 45 & 0.005\\
		0.05 & 256 & 50 & 40 & 0.01 \\
		0.1  & 64 & 20 & 16 & 0.025 \\
		0.2  & 64 & 20 & 16 & 0.025 \\
		0.3  & 64 & 20 & 16 & 0.025 \\
		0.3185  & 64 & 20 & 16 & 0.025 \\
		0.375  & 64 & 20 & 16 & 0.025 \\
		0.45  & 32 & 10 & 8 & 0.05 \\
		0.5  & 32 & 10 & 8 & 0.05 \\
		0.6  & 32 & 8 & 7 & 0.1
	\end{tabular}
	\label{tab1}
\end{table}
\clearpage
\newpage
\subsection{H$_2$ treatment in QED-CIS: Polaritonic states}
\label{appb}
The hydrogen molecule in a cavity, 
 treated  within the QED-CIS method, gives rise to
 polaritonic states, ${\Phi_{p}^\mathrm{QED}}$, 
 {\em cf.} Eq.(\ref{eq.qed_td_ci}).
 In Tab.~\ref{tab2}, we list lowest CIS and QED-CIS state energies 
 for H$_2$, for the latter for two different 
  choices of the
 cavity frequency, $\omega_c$.

\begin{table}[h!]
	\renewcommand{\baselinestretch}{1.}
	\caption{Lowest ten eigenstates 
		of H$_2$ obtained from CIS and QED-CIS calculations 
		(for $\omega_c=0.057$ or $\omega_c=0.467$, $g_c=0.01$ and $N_p=20$) 
		as described in the text (all in atomic units).}
	\begin{tabular}{c | c | p{2.4cm} | p{2.4cm}}
		state $i/p$   & $E_i$ (CIS)  & $E_p$ (QED-CIS, $\omega_c=0.057$) & $E_p$ (QED-CIS, $\omega_c=0.467$) \\
		\hline
\renewcommand{\baselinestretch}{1.}
		0 & -1.13314 & -1.12958 & -1.1327\\
		1 & -0.665451 & -1.07369 & -0.674301 \\
		2 & -0.655595 & -1.01746 & -0.654254\\ 
		3 & -0.652122 & -0.961236 & -0.653525 \\
		4 & -0.652122 & -0.90502 & -0.650543\\
		5 & -0.594238 & -0.848808 & -0.650543\\
		6 & -0.593195 & -0.792599 & -0.592787 \\
		7 & -0.592898 & -0.736393 & -0.591978\\
		8 & -0.592898 & -0.680192 & -0.591978\\
		9 & -0.591995 & -0.650749 & -0.590026
	\end{tabular}
	\label{tab2}
\end{table}

%
In the following figure \ref{fig:polaritons},
 the lowest QED-CIS eigenstates are given with their energy and decomposition into zero-order states, $|i,n\rangle=\ket{\Phi_i}\ket{\psi_n}$, for the two different choices of the
 cavity frequency. ($\ket{\Phi_i}$ indicates field- and cavity-free CIS states, and 
 $\ket{\psi_n}$ the eigenstates of the cavity Hamiltonian, $\hat{H}^c$.)

\begin{figure}[h!]
\begin{tabular}{cc}
a) & b) \\
\includegraphics[width=0.5\linewidth]{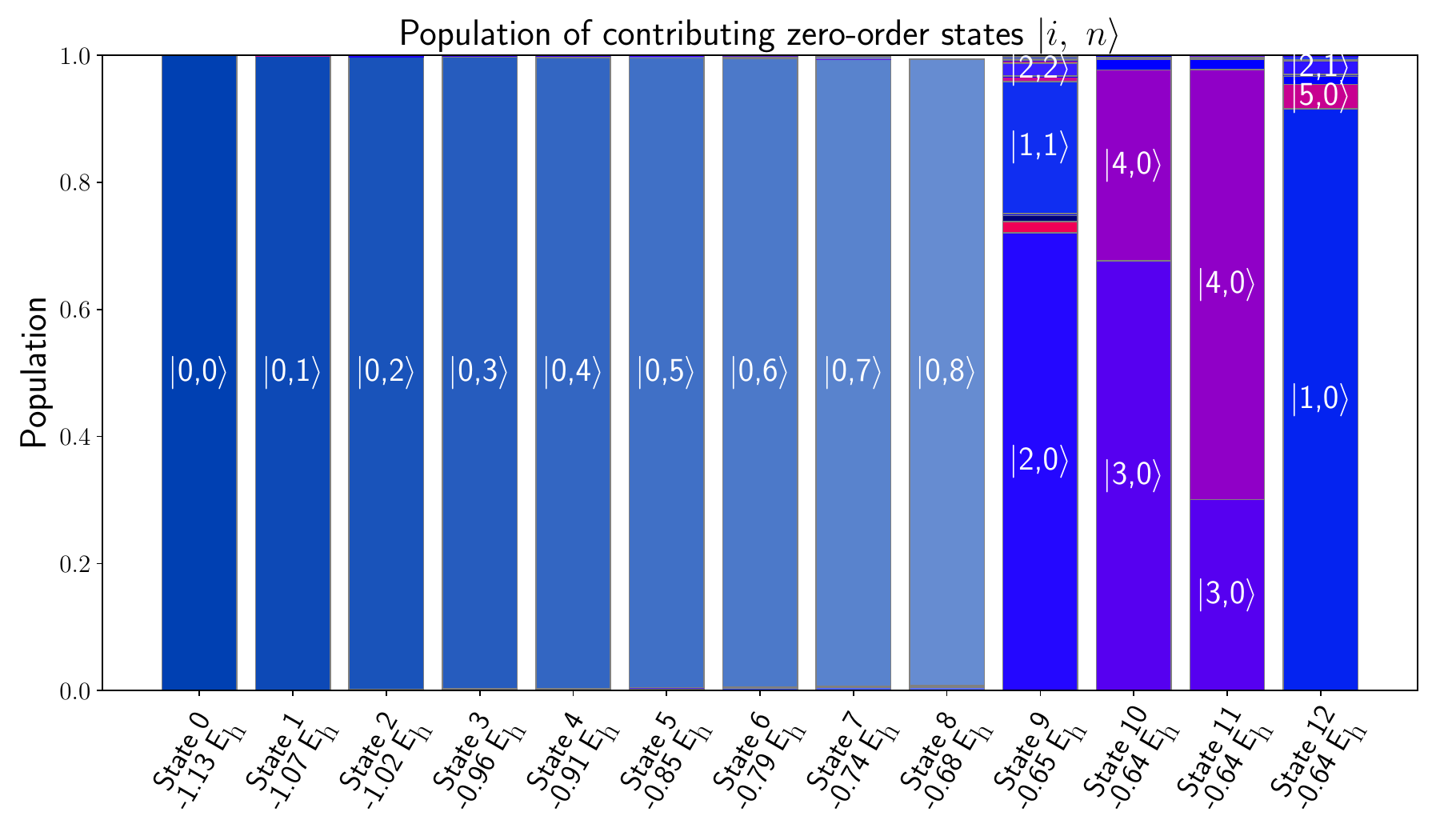}
 & \includegraphics[width=0.5\linewidth]{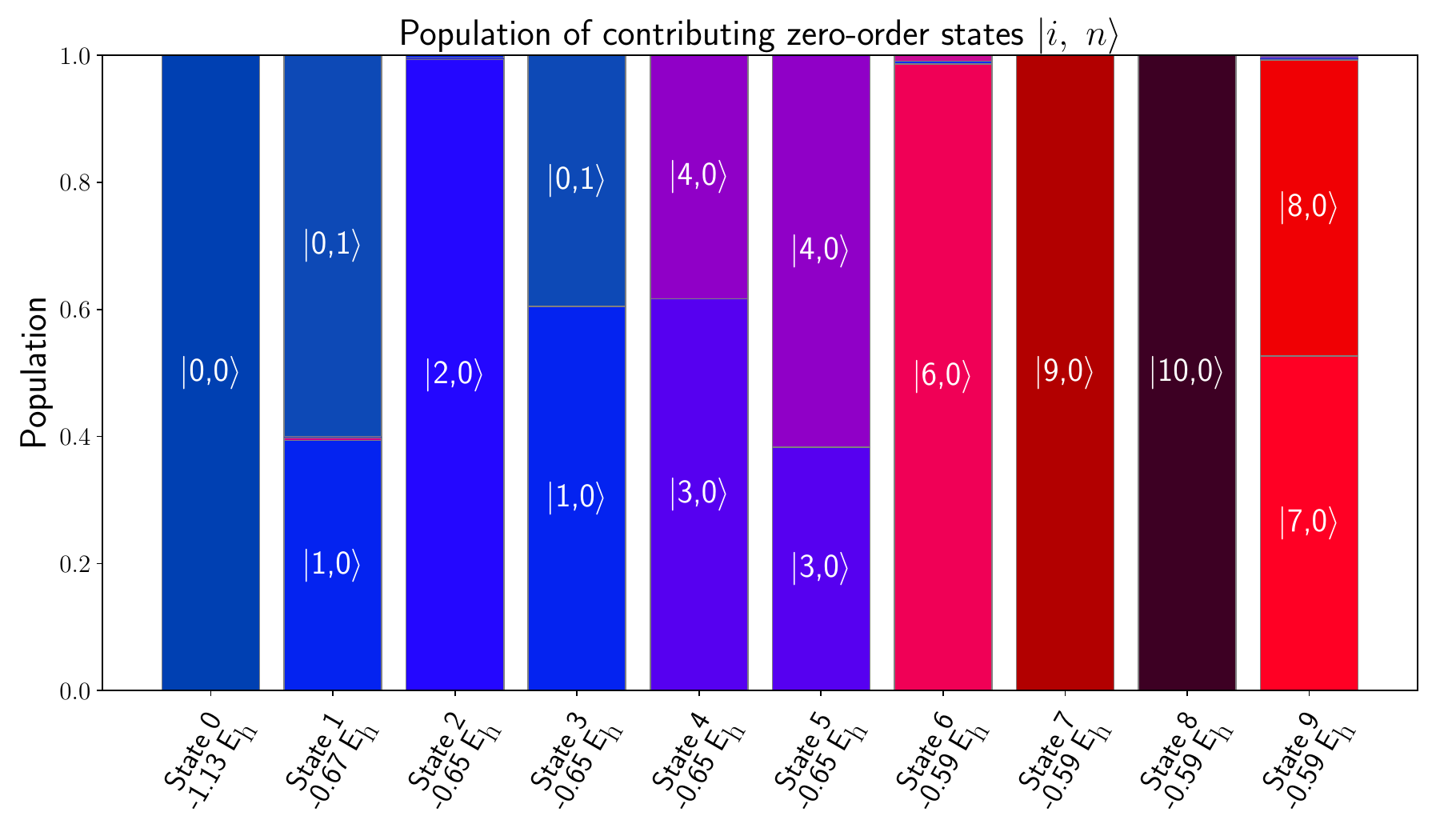}
\end{tabular}
\renewcommand{\baselinestretch}{1.}
	\caption{Decomposition of the lowest QED-CIS eigenstates of H$_2$ in a cavity, $\ket{\Phi_{p}^\mathrm{QED}}$, into zero-order product states $|i,n\rangle = \ket{\Phi_i}\ket{\psi_n}$.  The size of the sub-boxes in the individual 
 bars of the figure are proportional to the 
 coefficients squared $|D_{p,in}^\mathrm{QED}|^2$ of the corresponding zero-order product states as defined 
 in Eq.(\ref{eq.qed_td_ci}). Selected, dominating $|i,n\rangle$'s are indicated.
Color code: Blue to red indicates higher electronic excitation; higher white content
 indicates higher photonic excitation.
	 a) For the case, $\omega_c=0.057$, resonant to the laser frequency, $\omega_0$, with $N_p=20$ and $g_c=0.01$.
 	This results in lower-lying states, incrementally increasing in energy with the photon quanta, mainly consisting of the single photon excitation of the ground state, followed up by states with polaritonic character.
	 b) In the case, $\omega_c=0.468$, resonant to the first electronic transition, with $N_p=20$ and $g_c=0.01$. This results in an upper and lower polariton state (states 2 and 4), around the (formerly) degenerate state 3. 
	The other (low-energy) states do not show a strong polaritonic character.
}\label{fig:polaritons}
\end{figure}
\clearpage
\newpage
\subsection{H$_2$ treatment in QED-TD-CIS: Time evolution}
\label{appc}

The following two figures show the time-evolution of 
 QED-CIS state populations for H$_2$, 
 during excitation with a 800 nm laser pulse, 
 when the molecule was placed in a cavities of two different 
 cavity frequencies, $\omega_c$.

\begin{figure}
\begin{tabular}{cc}
 a) & b) \\
              \includegraphics[width=.5\linewidth]{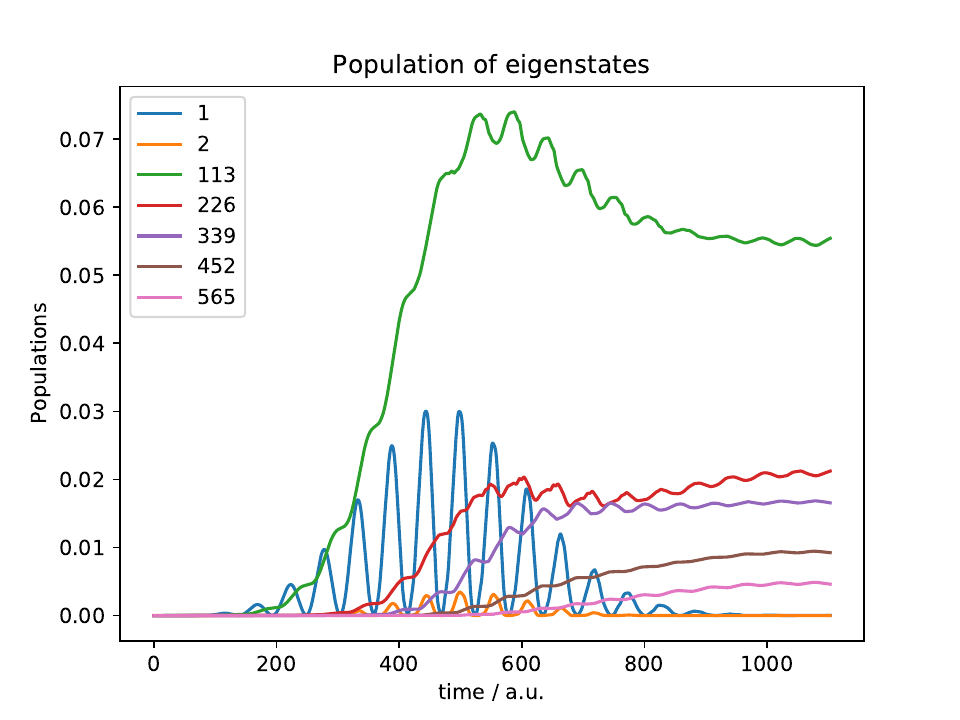}
             &   \includegraphics[width=.5\linewidth]{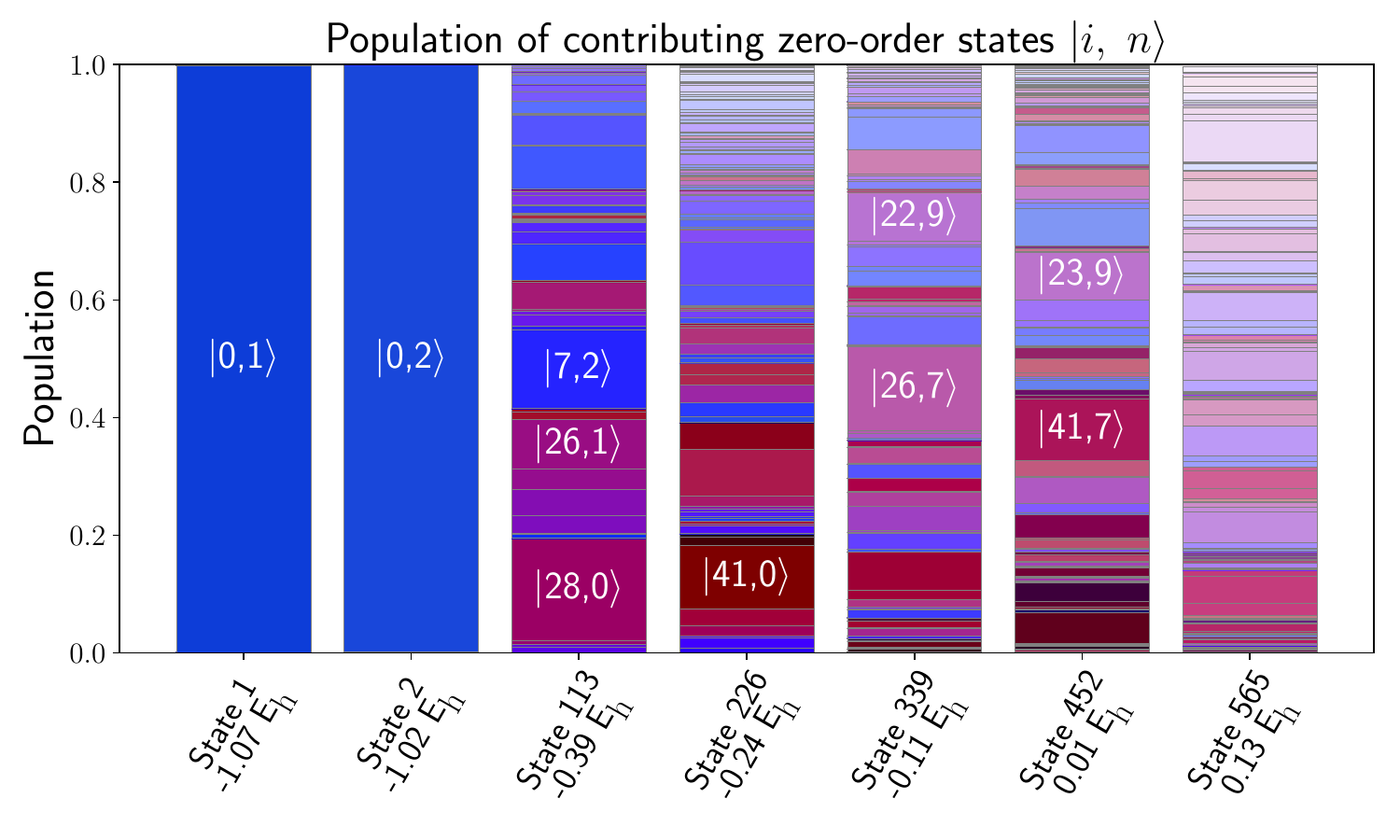}
\end{tabular}
\renewcommand{\baselinestretch}{1.}
                \caption{H$_2$ molecule calculated with QED-CIS: a) Time-dependent population of most populated, excited QED-CIS eigenstates for the propagation of H$_2$ using $N_p=20$, $g_c=0.01$ and ${\omega_c=0.057}$, \textit{i.e.}, resonant to an exciting 800~nm, 10-cycle laser pulse.
                b) Contributing zero-order state decomposition to the selected states under the same parameters. Same color coding as before; selected, dominating zero-order states 
 are highlighted.
                We see a lasting excitation into higher electronic states, which have a complex polaritonic composition.
                }
\label{t1}
        \end{figure}

        \begin{figure}
\begin{tabular}{cc}
 a) & b) \\
\includegraphics[width=.5\linewidth]{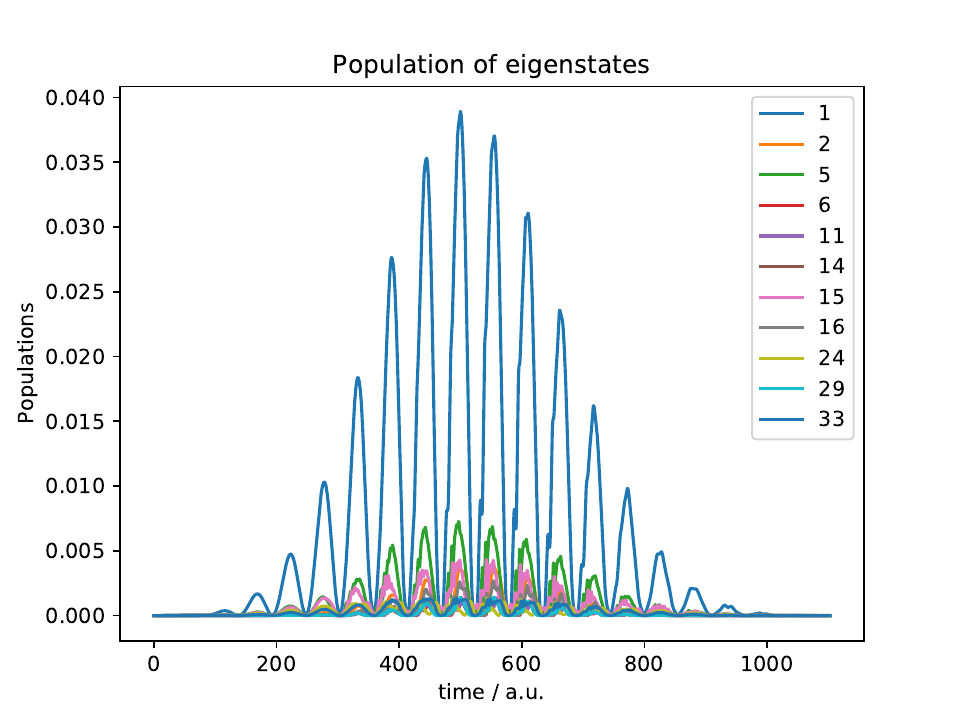}
        \includegraphics[width=.5\linewidth]{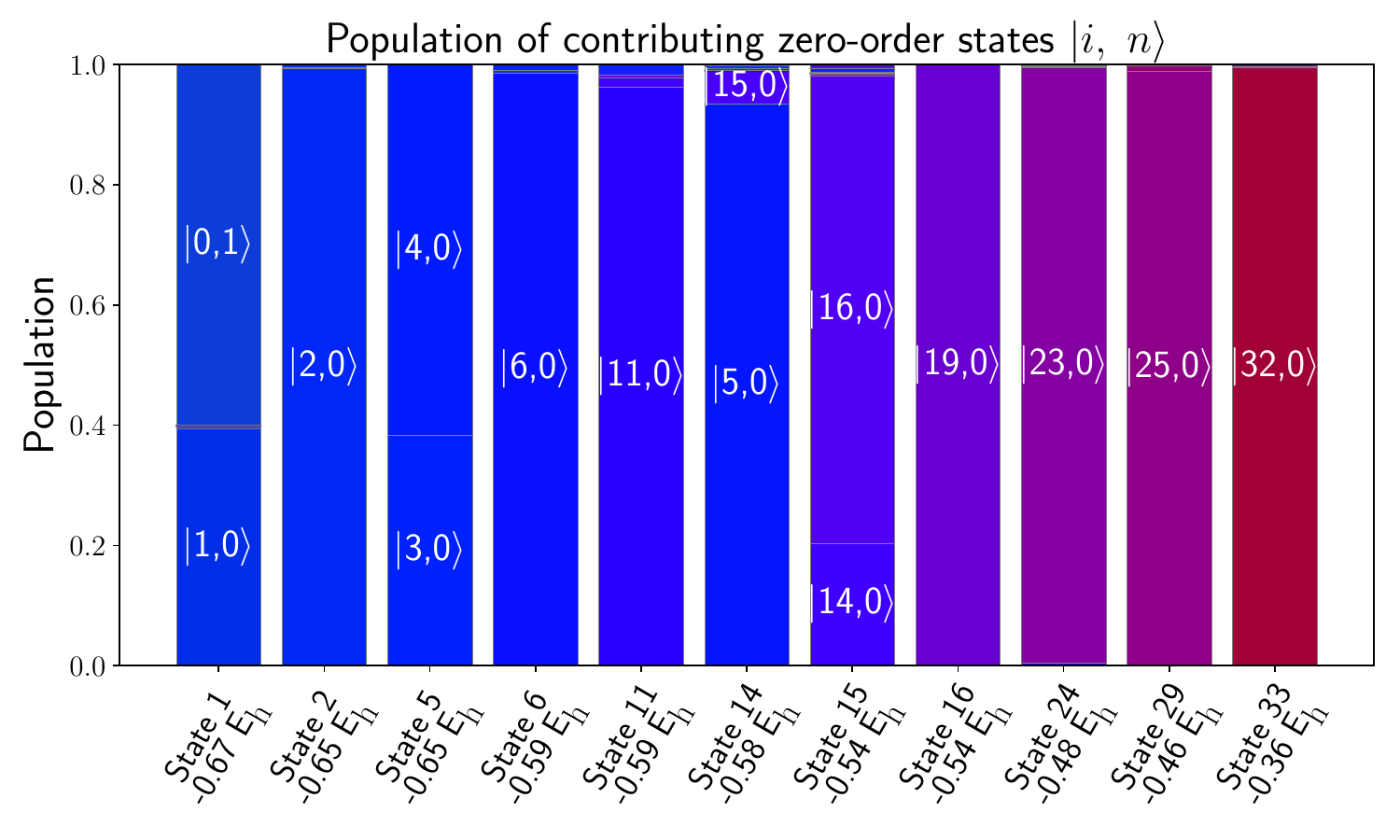}
\end{tabular}
\renewcommand{\baselinestretch}{1.}
        \caption{H$_2$ molecule calculated with QED-CIS: (a) Time-dependent population of most populated, excited QED-CIS eigenstates for the propagation of H$_2$ using $N_p=20$, $g_c=0.01$ and ${\omega_c=0.467}$, \textit{i.e.}, resonant to the first electronic transition. It gets excited with a 800~nm, 10-cycle laser pulse.
                b) Contributing zero-order state decomposition to the selected states under the same parameters.
  Same color coding as before; selected, dominating zero-order states
 are highlighted.
                We see a non-lasting excitation into higher electronic states, which are mostly purely electronic excitations.}
\label{t2}        
\end{figure}

\clearpage
\bibliography{literature}

\begin{thebibliography}{40}%
\makeatletter
\providecommand \@ifxundefined [1]{%
 \@ifx{#1\undefined}
}%
\providecommand \@ifnum [1]{%
 \ifnum #1\expandafter \@firstoftwo
 \else \expandafter \@secondoftwo
 \fi
}%
\providecommand \@ifx [1]{%
 \ifx #1\expandafter \@firstoftwo
 \else \expandafter \@secondoftwo
 \fi
}%
\providecommand \natexlab [1]{#1}%
\providecommand \enquote  [1]{``#1''}%
\providecommand \bibnamefont  [1]{#1}%
\providecommand \bibfnamefont [1]{#1}%
\providecommand \citenamefont [1]{#1}%
\providecommand \href@noop [0]{\@secondoftwo}%
\providecommand \href [0]{\begingroup \@sanitize@url \@href}%
\providecommand \@href[1]{\@@startlink{#1}\@@href}%
\providecommand \@@href[1]{\endgroup#1\@@endlink}%
\providecommand \@sanitize@url [0]{\catcode `\\12\catcode `\$12\catcode
  `\&12\catcode `\#12\catcode `\^12\catcode `\_12\catcode `\%12\relax}%
\providecommand \@@startlink[1]{}%
\providecommand \@@endlink[0]{}%
\providecommand \url  [0]{\begingroup\@sanitize@url \@url }%
\providecommand \@url [1]{\endgroup\@href {#1}{\urlprefix }}%
\providecommand \urlprefix  [0]{URL }%
\providecommand \Eprint [0]{\href }%
\providecommand \doibase [0]{https://doi.org/}%
\providecommand \selectlanguage [0]{\@gobble}%
\providecommand \bibinfo  [0]{\@secondoftwo}%
\providecommand \bibfield  [0]{\@secondoftwo}%
\providecommand \translation [1]{[#1]}%
\providecommand \BibitemOpen [0]{}%
\providecommand \bibitemStop [0]{}%
\providecommand \bibitemNoStop [0]{.\EOS\space}%
\providecommand \EOS [0]{\spacefactor3000\relax}%
\providecommand \BibitemShut  [1]{\csname bibitem#1\endcsname}%
\let\auto@bib@innerbib\@empty
\bibitem [{\citenamefont {Corkum}\ and\ \citenamefont
  {Krausz}(2007)}]{Corkum2007}%
  \BibitemOpen
  \bibfield  {author} {\bibinfo {author} {\bibfnamefont {P.~B.}\ \bibnamefont
  {Corkum}}\ and\ \bibinfo {author} {\bibfnamefont {F.}~\bibnamefont
  {Krausz}},\ }\href@noop {} {\bibfield  {journal} {\bibinfo  {journal} {Nat.
  Phys.}\ }\textbf {\bibinfo {volume} {3}},\ \bibinfo {pages} {381} (\bibinfo
  {year} {2007})}\BibitemShut {NoStop}%
\bibitem [{\citenamefont {Sansone}, \citenamefont {Kelkensberg},\ and\
  \citenamefont {P\'erez-Torres}(2010)}]{Sansone2010}%
  \BibitemOpen
  \bibfield  {author} {\bibinfo {author} {\bibfnamefont {G.}~\bibnamefont
  {Sansone}}, \bibinfo {author} {\bibfnamefont {F.}~\bibnamefont
  {Kelkensberg}},\ and\ \bibinfo {author} {\bibfnamefont {J.}~\bibnamefont
  {P\'erez-Torres}},\ }\href@noop {} {\bibfield  {journal} {\bibinfo  {journal}
  {Nat.}\ ,\ \bibinfo {pages} {763}} (\bibinfo {year} {2010})}\BibitemShut
  {NoStop}%
\bibitem [{\citenamefont {Nisoli}\ \emph {et~al.}(2017)\citenamefont {Nisoli},
  \citenamefont {Decleva}, \citenamefont {Calegari}, \citenamefont {Palacios},\
  and\ \citenamefont {Mart\'in}}]{Nisolini2017}%
  \BibitemOpen
  \bibfield  {author} {\bibinfo {author} {\bibfnamefont {M.}~\bibnamefont
  {Nisoli}}, \bibinfo {author} {\bibfnamefont {P.}~\bibnamefont {Decleva}},
  \bibinfo {author} {\bibfnamefont {F.}~\bibnamefont {Calegari}}, \bibinfo
  {author} {\bibfnamefont {A.}~\bibnamefont {Palacios}},\ and\ \bibinfo
  {author} {\bibfnamefont {F.}~\bibnamefont {Mart\'in}},\ }\href@noop {}
  {\bibfield  {journal} {\bibinfo  {journal} {Chem. Rev.}\ }\textbf {\bibinfo
  {volume} {117}},\ \bibinfo {pages} {10760} (\bibinfo {year}
  {2017})}\BibitemShut {NoStop}%
\bibitem [{\citenamefont {Gaumnitz}\ \emph {et~al.}(2017)\citenamefont
  {Gaumnitz}, \citenamefont {Jain}, \citenamefont {Pertot}, \citenamefont
  {Huppert}, \citenamefont {Jordan}, \citenamefont {Ardana-Lamas},\ and\
  \citenamefont {W\"orner}}]{Gaumnitz2017}%
  \BibitemOpen
  \bibfield  {author} {\bibinfo {author} {\bibfnamefont {T.}~\bibnamefont
  {Gaumnitz}}, \bibinfo {author} {\bibfnamefont {A.}~\bibnamefont {Jain}},
  \bibinfo {author} {\bibfnamefont {Y.}~\bibnamefont {Pertot}}, \bibinfo
  {author} {\bibfnamefont {M.}~\bibnamefont {Huppert}}, \bibinfo {author}
  {\bibfnamefont {I.}~\bibnamefont {Jordan}}, \bibinfo {author} {\bibfnamefont
  {F.}~\bibnamefont {Ardana-Lamas}},\ and\ \bibinfo {author} {\bibfnamefont
  {H.}~\bibnamefont {W\"orner}},\ }\href@noop {} {\bibfield  {journal}
  {\bibinfo  {journal} {Opt. Express}\ }\textbf {\bibinfo {volume} {25}},\
  \bibinfo {pages} {27506} (\bibinfo {year} {2017})}\BibitemShut {NoStop}%
\bibitem [{\citenamefont {Corkum}(1993)}]{Corkum1993}%
  \BibitemOpen
  \bibfield  {author} {\bibinfo {author} {\bibfnamefont {P.~B.}\ \bibnamefont
  {Corkum}},\ }\href@noop {} {\bibfield  {journal} {\bibinfo  {journal} {Phys.
  Rev. Lett.}\ }\textbf {\bibinfo {volume} {71}},\ \bibinfo {pages} {1994}
  (\bibinfo {year} {1993})}\BibitemShut {NoStop}%
\bibitem [{\citenamefont {Weidman}\ \emph {et~al.}(2024)\citenamefont
  {Weidman}, \citenamefont {Dadgar}, \citenamefont {Stewart}, \citenamefont
  {Peyton}, \citenamefont {Ulusoy},\ and\ \citenamefont {Wilson}}]{qedtdci}%
  \BibitemOpen
  \bibfield  {author} {\bibinfo {author} {\bibfnamefont {J.~D.}\ \bibnamefont
  {Weidman}}, \bibinfo {author} {\bibfnamefont {M.~S.}\ \bibnamefont {Dadgar}},
  \bibinfo {author} {\bibfnamefont {Z.~J.}\ \bibnamefont {Stewart}}, \bibinfo
  {author} {\bibfnamefont {B.~G.}\ \bibnamefont {Peyton}}, \bibinfo {author}
  {\bibfnamefont {I.~S.}\ \bibnamefont {Ulusoy}},\ and\ \bibinfo {author}
  {\bibfnamefont {A.~K.}\ \bibnamefont {Wilson}},\ }\href@noop {} {\bibfield
  {journal} {\bibinfo  {journal} {J. Chem. Phys.}\ }\textbf {\bibinfo {volume}
  {160}},\ \bibinfo {pages} {094111} (\bibinfo {year} {2024})}\BibitemShut
  {NoStop}%
\bibitem [{\citenamefont {Gohle}\ \emph {et~al.}(2005)\citenamefont {Gohle},
  \citenamefont {Udem}, \citenamefont {Herrmann}, \citenamefont
  {Rauschenberger}, \citenamefont {Holzwarth}, \citenamefont {Schuessler},
  \citenamefont {Krausz},\ and\ \citenamefont {H{\"a}nsch}}]{Jour2005}%
  \BibitemOpen
  \bibfield  {author} {\bibinfo {author} {\bibfnamefont {C.}~\bibnamefont
  {Gohle}}, \bibinfo {author} {\bibfnamefont {T.}~\bibnamefont {Udem}},
  \bibinfo {author} {\bibfnamefont {M.}~\bibnamefont {Herrmann}}, \bibinfo
  {author} {\bibfnamefont {J.}~\bibnamefont {Rauschenberger}}, \bibinfo
  {author} {\bibfnamefont {R.}~\bibnamefont {Holzwarth}}, \bibinfo {author}
  {\bibfnamefont {H.~A.}\ \bibnamefont {Schuessler}}, \bibinfo {author}
  {\bibfnamefont {F.}~\bibnamefont {Krausz}},\ and\ \bibinfo {author}
  {\bibfnamefont {T.~W.}\ \bibnamefont {H{\"a}nsch}},\ }\href@noop {}
  {\bibfield  {journal} {\bibinfo  {journal} {Nat.}\ }\textbf {\bibinfo
  {volume} {436}},\ \bibinfo {pages} {234} (\bibinfo {year}
  {2005})}\BibitemShut {NoStop}%
\bibitem [{\citenamefont {Jones}\ \emph {et~al.}(2005)\citenamefont {Jones},
  \citenamefont {Moll}, \citenamefont {Thorpe},\ and\ \citenamefont
  {Ye}}]{Jones2005}%
  \BibitemOpen
  \bibfield  {author} {\bibinfo {author} {\bibfnamefont {R.~J.}\ \bibnamefont
  {Jones}}, \bibinfo {author} {\bibfnamefont {K.~D.}\ \bibnamefont {Moll}},
  \bibinfo {author} {\bibfnamefont {M.~J.}\ \bibnamefont {Thorpe}},\ and\
  \bibinfo {author} {\bibfnamefont {J.}~\bibnamefont {Ye}},\ }\href@noop {}
  {\bibfield  {journal} {\bibinfo  {journal} {Phys. Rev. Lett.}\ }\textbf
  {\bibinfo {volume} {94}},\ \bibinfo {pages} {193201} (\bibinfo {year}
  {2005})}\BibitemShut {NoStop}%
\bibitem [{\citenamefont {H\"ogner}\ \emph {et~al.}(2019)\citenamefont
  {H\"ogner}, \citenamefont {Saule}, \citenamefont {Heinrich}, \citenamefont
  {Lilienfein}, \citenamefont {Esser}, \citenamefont {Trubetskov},
  \citenamefont {Pervak},\ and\ \citenamefont {Pupeza}}]{Hogner2019}%
  \BibitemOpen
  \bibfield  {author} {\bibinfo {author} {\bibfnamefont {M.}~\bibnamefont
  {H\"ogner}}, \bibinfo {author} {\bibfnamefont {T.}~\bibnamefont {Saule}},
  \bibinfo {author} {\bibfnamefont {S.}~\bibnamefont {Heinrich}}, \bibinfo
  {author} {\bibfnamefont {N.}~\bibnamefont {Lilienfein}}, \bibinfo {author}
  {\bibfnamefont {D.}~\bibnamefont {Esser}}, \bibinfo {author} {\bibfnamefont
  {M.}~\bibnamefont {Trubetskov}}, \bibinfo {author} {\bibfnamefont
  {V.}~\bibnamefont {Pervak}},\ and\ \bibinfo {author} {\bibfnamefont
  {I.}~\bibnamefont {Pupeza}},\ }\href@noop {} {\bibfield  {journal} {\bibinfo
  {journal} {Opt. Express}\ }\textbf {\bibinfo {volume} {27}},\ \bibinfo
  {pages} {19675} (\bibinfo {year} {2019})}\BibitemShut {NoStop}%
\bibitem [{\citenamefont {Kim}\ \emph {et~al.}(2008)\citenamefont {Kim},
  \citenamefont {Jin}, \citenamefont {Kim}, \citenamefont {Park}, \citenamefont
  {Kim},\ and\ \citenamefont {Kim}}]{Kim2008}%
  \BibitemOpen
  \bibfield  {author} {\bibinfo {author} {\bibfnamefont {S.}~\bibnamefont
  {Kim}}, \bibinfo {author} {\bibfnamefont {J.}~\bibnamefont {Jin}}, \bibinfo
  {author} {\bibfnamefont {Y.-J.}\ \bibnamefont {Kim}}, \bibinfo {author}
  {\bibfnamefont {I.-Y.}\ \bibnamefont {Park}}, \bibinfo {author}
  {\bibfnamefont {Y.}~\bibnamefont {Kim}},\ and\ \bibinfo {author}
  {\bibfnamefont {S.-W.}\ \bibnamefont {Kim}},\ }\href@noop {} {\bibfield
  {journal} {\bibinfo  {journal} {Nat.}\ }\textbf {\bibinfo {volume} {453}},\
  \bibinfo {pages} {757} (\bibinfo {year} {2008})}\BibitemShut {NoStop}%
\bibitem [{\citenamefont {Park}\ \emph {et~al.}(2011)\citenamefont {Park},
  \citenamefont {Kim}, \citenamefont {Choi}, \citenamefont {Lee}, \citenamefont
  {Kim}, \citenamefont {Kling}, \citenamefont {Stockman},\ and\ \citenamefont
  {Kim}}]{Park2011}%
  \BibitemOpen
  \bibfield  {author} {\bibinfo {author} {\bibfnamefont {I.-Y.}\ \bibnamefont
  {Park}}, \bibinfo {author} {\bibfnamefont {S.}~\bibnamefont {Kim}}, \bibinfo
  {author} {\bibfnamefont {J.}~\bibnamefont {Choi}}, \bibinfo {author}
  {\bibfnamefont {D.-H.}\ \bibnamefont {Lee}}, \bibinfo {author} {\bibfnamefont
  {Y.-J.}\ \bibnamefont {Kim}}, \bibinfo {author} {\bibfnamefont {M.~F.}\
  \bibnamefont {Kling}}, \bibinfo {author} {\bibfnamefont {M.~I.}\ \bibnamefont
  {Stockman}},\ and\ \bibinfo {author} {\bibfnamefont {S.-W.}\ \bibnamefont
  {Kim}},\ }\href@noop {} {\bibfield  {journal} {\bibinfo  {journal} {Nat.
  Photonics}\ }\textbf {\bibinfo {volume} {5}},\ \bibinfo {pages} {677}
  (\bibinfo {year} {2011})}\BibitemShut {NoStop}%
\bibitem [{\citenamefont {Ebadian}\ and\ \citenamefont
  {Mohebbi}(2017)}]{Ebadian2017}%
  \BibitemOpen
  \bibfield  {author} {\bibinfo {author} {\bibfnamefont {H.}~\bibnamefont
  {Ebadian}}\ and\ \bibinfo {author} {\bibfnamefont {M.}~\bibnamefont
  {Mohebbi}},\ }\href@noop {} {\bibfield  {journal} {\bibinfo  {journal} {Phys.
  Rev. A}\ }\textbf {\bibinfo {volume} {96}},\ \bibinfo {pages} {023415}
  (\bibinfo {year} {2017})}\BibitemShut {NoStop}%
\bibitem [{\citenamefont {Shahnavaz}\ and\ \citenamefont
  {Mohebbi}(2021)}]{Shahnavaz2021}%
  \BibitemOpen
  \bibfield  {author} {\bibinfo {author} {\bibfnamefont {N.}~\bibnamefont
  {Shahnavaz}}\ and\ \bibinfo {author} {\bibfnamefont {M.}~\bibnamefont
  {Mohebbi}},\ }\href@noop {} {\bibfield  {journal} {\bibinfo  {journal}
  {Plasmonics}\ }\textbf {\bibinfo {volume} {16}},\ \bibinfo {pages} {305}
  (\bibinfo {year} {2021})}\BibitemShut {NoStop}%
\bibitem [{\citenamefont {Gorlach}\ \emph {et~al.}(2023)\citenamefont
  {Gorlach}, \citenamefont {Tzur}, \citenamefont {Birk}, \citenamefont
  {Kr\"uger}, \citenamefont {Rivera}, \citenamefont {Cohen},\ and\
  \citenamefont {Kaminer}}]{Gorlach2023}%
  \BibitemOpen
  \bibfield  {author} {\bibinfo {author} {\bibfnamefont {A.}~\bibnamefont
  {Gorlach}}, \bibinfo {author} {\bibfnamefont {M.}~\bibnamefont {Tzur}},
  \bibinfo {author} {\bibfnamefont {M.}~\bibnamefont {Birk}}, \bibinfo {author}
  {\bibfnamefont {M.}~\bibnamefont {Kr\"uger}}, \bibinfo {author}
  {\bibfnamefont {N.}~\bibnamefont {Rivera}}, \bibinfo {author} {\bibfnamefont
  {O.}~\bibnamefont {Cohen}},\ and\ \bibinfo {author} {\bibfnamefont
  {I.}~\bibnamefont {Kaminer}},\ }\href@noop {} {\bibfield  {journal} {\bibinfo
   {journal} {Nat. Phys.}\ }\textbf {\bibinfo {volume} {19}},\ \bibinfo {pages}
  {1689} (\bibinfo {year} {2023})}\BibitemShut {NoStop}%
\bibitem [{\citenamefont {Aklilu}\ and\ \citenamefont
  {Varga}(2024)}]{Aklilu2024}%
  \BibitemOpen
  \bibfield  {author} {\bibinfo {author} {\bibfnamefont {Y.~S.}\ \bibnamefont
  {Aklilu}}\ and\ \bibinfo {author} {\bibfnamefont {K.}~\bibnamefont {Varga}},\
  }\href@noop {} {\bibfield  {journal} {\bibinfo  {journal} {Phys. Rev. A}\
  }\textbf {\bibinfo {volume} {110}},\ \bibinfo {pages} {043119} (\bibinfo
  {year} {2024})}\BibitemShut {NoStop}%
\bibitem [{\citenamefont {Haugland}\ \emph {et~al.}(2020)\citenamefont
  {Haugland}, \citenamefont {Ronca}, \citenamefont {Kj{\o}nstad}, \citenamefont
  {Rubio},\ and\ \citenamefont {Koch}}]{Haugland2020}%
  \BibitemOpen
  \bibfield  {author} {\bibinfo {author} {\bibfnamefont {T.}~\bibnamefont
  {Haugland}}, \bibinfo {author} {\bibfnamefont {E.}~\bibnamefont {Ronca}},
  \bibinfo {author} {\bibfnamefont {E.}~\bibnamefont {Kj{\o}nstad}}, \bibinfo
  {author} {\bibfnamefont {A.}~\bibnamefont {Rubio}},\ and\ \bibinfo {author}
  {\bibfnamefont {H.}~\bibnamefont {Koch}},\ }\href@noop {} {\bibfield
  {journal} {\bibinfo  {journal} {Phys. Rev. X}\ }\textbf {\bibinfo {volume}
  {10}},\ \bibinfo {pages} {041043} (\bibinfo {year} {2020})}\BibitemShut
  {NoStop}%
\bibitem [{\citenamefont {DePrince}(2021)}]{cavityionization}%
  \BibitemOpen
  \bibfield  {author} {\bibinfo {author} {\bibfnamefont {I.}~\bibnamefont
  {DePrince}, \bibfnamefont {A.~Eugene}},\ }\href@noop {} {\bibfield  {journal}
  {\bibinfo  {journal} {J. Chem. Phys.}\ }\textbf {\bibinfo {volume} {154}},\
  \bibinfo {pages} {094112} (\bibinfo {year} {2021})}\BibitemShut {NoStop}%
\bibitem [{\citenamefont {Tokatly}(2013)}]{Tokatly2013}%
  \BibitemOpen
  \bibfield  {author} {\bibinfo {author} {\bibfnamefont {I.~V.}\ \bibnamefont
  {Tokatly}},\ }\href@noop {} {\bibfield  {journal} {\bibinfo  {journal} {Phys.
  Rev. Lett.}\ }\textbf {\bibinfo {volume} {110}},\ \bibinfo {pages} {233001}
  (\bibinfo {year} {2013})}\BibitemShut {NoStop}%
\bibitem [{\citenamefont {Ruggenthaler}\ \emph {et~al.}(2014)\citenamefont
  {Ruggenthaler}, \citenamefont {Flick}, \citenamefont {Pellegrini},
  \citenamefont {Appel}, \citenamefont {Tokatly},\ and\ \citenamefont
  {Rubio}}]{Ruggenthaler2014}%
  \BibitemOpen
  \bibfield  {author} {\bibinfo {author} {\bibfnamefont {M.}~\bibnamefont
  {Ruggenthaler}}, \bibinfo {author} {\bibfnamefont {J.}~\bibnamefont {Flick}},
  \bibinfo {author} {\bibfnamefont {C.}~\bibnamefont {Pellegrini}}, \bibinfo
  {author} {\bibfnamefont {H.}~\bibnamefont {Appel}}, \bibinfo {author}
  {\bibfnamefont {I.~V.}\ \bibnamefont {Tokatly}},\ and\ \bibinfo {author}
  {\bibfnamefont {A.}~\bibnamefont {Rubio}},\ }\href@noop {} {\bibfield
  {journal} {\bibinfo  {journal} {Phys. Rev. A}\ }\textbf {\bibinfo {volume}
  {90}},\ \bibinfo {pages} {012508} (\bibinfo {year} {2014})}\BibitemShut
  {NoStop}%
\bibitem [{\citenamefont {Flick}\ \emph
  {et~al.}(2017{\natexlab{a}})\citenamefont {Flick}, \citenamefont
  {Ruggenthaler}, \citenamefont {Appel},\ and\ \citenamefont
  {Rubio}}]{Flick2017atoms}%
  \BibitemOpen
  \bibfield  {author} {\bibinfo {author} {\bibfnamefont {J.}~\bibnamefont
  {Flick}}, \bibinfo {author} {\bibfnamefont {M.}~\bibnamefont {Ruggenthaler}},
  \bibinfo {author} {\bibfnamefont {H.}~\bibnamefont {Appel}},\ and\ \bibinfo
  {author} {\bibfnamefont {A.}~\bibnamefont {Rubio}},\ }\href@noop {}
  {\bibfield  {journal} {\bibinfo  {journal} {Proc. Natl. Acad. Sci. U.S.A.}\
  }\textbf {\bibinfo {volume} {114}},\ \bibinfo {pages} {3026} (\bibinfo {year}
  {2017}{\natexlab{a}})}\BibitemShut {NoStop}%
\bibitem [{\citenamefont {Haugland}\ \emph {et~al.}(2021)\citenamefont
  {Haugland}, \citenamefont {Sch\"afer}, \citenamefont {Ronca}, \citenamefont
  {Rubio},\ and\ \citenamefont {Koch}}]{Haugland2021}%
  \BibitemOpen
  \bibfield  {author} {\bibinfo {author} {\bibfnamefont {T.}~\bibnamefont
  {Haugland}}, \bibinfo {author} {\bibfnamefont {C.}~\bibnamefont {Sch\"afer}},
  \bibinfo {author} {\bibfnamefont {E.}~\bibnamefont {Ronca}}, \bibinfo
  {author} {\bibfnamefont {A.}~\bibnamefont {Rubio}},\ and\ \bibinfo {author}
  {\bibfnamefont {H.}~\bibnamefont {Koch}},\ }\href@noop {} {\bibfield
  {journal} {\bibinfo  {journal} {J. Chem. Phys.}\ }\textbf {\bibinfo {volume}
  {154}},\ \bibinfo {pages} {094113} (\bibinfo {year} {2021})}\BibitemShut
  {NoStop}%
\bibitem [{\citenamefont {Bedurke}\ \emph {et~al.}(2019)\citenamefont
  {Bedurke}, \citenamefont {Klamroth}, \citenamefont {Krause},\ and\
  \citenamefont {Saalfrank}}]{bedurke2019discriminating}%
  \BibitemOpen
  \bibfield  {author} {\bibinfo {author} {\bibfnamefont {F.}~\bibnamefont
  {Bedurke}}, \bibinfo {author} {\bibfnamefont {T.}~\bibnamefont {Klamroth}},
  \bibinfo {author} {\bibfnamefont {P.}~\bibnamefont {Krause}},\ and\ \bibinfo
  {author} {\bibfnamefont {P.}~\bibnamefont {Saalfrank}},\ }\href@noop {}
  {\bibfield  {journal} {\bibinfo  {journal} {J. Chem. Phys.}\ }\textbf
  {\bibinfo {volume} {150}} (\bibinfo {year} {2019})}\BibitemShut {NoStop}%
\bibitem [{\citenamefont {Saalfrank}\ \emph {et~al.}(2020)\citenamefont
  {Saalfrank}, \citenamefont {Bedurke}, \citenamefont {Heide}, \citenamefont
  {Klamroth}, \citenamefont {Klinkusch}, \citenamefont {Krause}, \citenamefont
  {Nest},\ and\ \citenamefont {Tremblay}}]{saalfrank2020molecular}%
  \BibitemOpen
  \bibfield  {author} {\bibinfo {author} {\bibfnamefont {P.}~\bibnamefont
  {Saalfrank}}, \bibinfo {author} {\bibfnamefont {F.}~\bibnamefont {Bedurke}},
  \bibinfo {author} {\bibfnamefont {C.}~\bibnamefont {Heide}}, \bibinfo
  {author} {\bibfnamefont {T.}~\bibnamefont {Klamroth}}, \bibinfo {author}
  {\bibfnamefont {S.}~\bibnamefont {Klinkusch}}, \bibinfo {author}
  {\bibfnamefont {P.}~\bibnamefont {Krause}}, \bibinfo {author} {\bibfnamefont
  {M.}~\bibnamefont {Nest}},\ and\ \bibinfo {author} {\bibfnamefont {J.~C.}\
  \bibnamefont {Tremblay}}\ }(\bibinfo  {publisher} {Elsevier},\ \bibinfo
  {year} {2020})\ p.~\bibinfo {pages} {15}\BibitemShut {NoStop}%
\bibitem [{\citenamefont {Bedurke}, \citenamefont {Klamroth},\ and\
  \citenamefont {Saalfrank}(2021)}]{bedurke2021many}%
  \BibitemOpen
  \bibfield  {author} {\bibinfo {author} {\bibfnamefont {F.}~\bibnamefont
  {Bedurke}}, \bibinfo {author} {\bibfnamefont {T.}~\bibnamefont {Klamroth}},\
  and\ \bibinfo {author} {\bibfnamefont {P.}~\bibnamefont {Saalfrank}},\
  }\href@noop {} {\bibfield  {journal} {\bibinfo  {journal} {Phys. Chem. Chem.
  Phys.}\ }\textbf {\bibinfo {volume} {23}},\ \bibinfo {pages} {13544}
  (\bibinfo {year} {2021})}\BibitemShut {NoStop}%
\bibitem [{\citenamefont {Flick}\ \emph
  {et~al.}(2017{\natexlab{b}})\citenamefont {Flick}, \citenamefont {Appel},
  \citenamefont {Ruggenthaler},\ and\ \citenamefont {Rubio}}]{Flick2017}%
  \BibitemOpen
  \bibfield  {author} {\bibinfo {author} {\bibfnamefont {J.}~\bibnamefont
  {Flick}}, \bibinfo {author} {\bibfnamefont {H.}~\bibnamefont {Appel}},
  \bibinfo {author} {\bibfnamefont {M.}~\bibnamefont {Ruggenthaler}},\ and\
  \bibinfo {author} {\bibfnamefont {A.}~\bibnamefont {Rubio}},\ }\href@noop {}
  {\bibfield  {journal} {\bibinfo  {journal} {J. Chem. Theory Comput.}\
  }\textbf {\bibinfo {volume} {13}},\ \bibinfo {pages} {1616} (\bibinfo {year}
  {2017}{\natexlab{b}})}\BibitemShut {NoStop}%
\bibitem [{\citenamefont {Sch\"afer}, \citenamefont {Ruggenthaler},\ and\
  \citenamefont {Rubio}(2018)}]{Schaefer2018}%
  \BibitemOpen
  \bibfield  {author} {\bibinfo {author} {\bibfnamefont {C.}~\bibnamefont
  {Sch\"afer}}, \bibinfo {author} {\bibfnamefont {M.}~\bibnamefont
  {Ruggenthaler}},\ and\ \bibinfo {author} {\bibfnamefont {A.}~\bibnamefont
  {Rubio}},\ }\href@noop {} {\bibfield  {journal} {\bibinfo  {journal} {Phys.
  Rev. A}\ }\textbf {\bibinfo {volume} {98}},\ \bibinfo {pages} {043801}
  (\bibinfo {year} {2018})}\BibitemShut {NoStop}%
\bibitem [{\citenamefont {Li}, \citenamefont {Mandal},\ and\ \citenamefont
  {Huo}(2021)}]{Li2021}%
  \BibitemOpen
  \bibfield  {author} {\bibinfo {author} {\bibfnamefont {X.}~\bibnamefont
  {Li}}, \bibinfo {author} {\bibfnamefont {A.}~\bibnamefont {Mandal}},\ and\
  \bibinfo {author} {\bibfnamefont {P.}~\bibnamefont {Huo}},\ }\href@noop {}
  {\bibfield  {journal} {\bibinfo  {journal} {Nat. commun.}\ }\textbf {\bibinfo
  {volume} {12}},\ \bibinfo {pages} {1315} (\bibinfo {year}
  {2021})}\BibitemShut {NoStop}%
\bibitem [{\citenamefont {Fischer}\ and\ \citenamefont
  {Saalfrank}(2023{\natexlab{a}})}]{Fischer2023cbo}%
  \BibitemOpen
  \bibfield  {author} {\bibinfo {author} {\bibfnamefont {E.~W.}\ \bibnamefont
  {Fischer}}\ and\ \bibinfo {author} {\bibfnamefont {P.}~\bibnamefont
  {Saalfrank}},\ }\href@noop {} {\bibfield  {journal} {\bibinfo  {journal} {J.
  Chem. Theory Comput.}\ }\textbf {\bibinfo {volume} {19}},\ \bibinfo {pages}
  {7215} (\bibinfo {year} {2023}{\natexlab{a}})}\BibitemShut {NoStop}%
\bibitem [{\citenamefont {Fischer}\ and\ \citenamefont
  {Saalfrank}(2023{\natexlab{b}})}]{Fischer2023}%
  \BibitemOpen
  \bibfield  {author} {\bibinfo {author} {\bibfnamefont {E.~W.}\ \bibnamefont
  {Fischer}}\ and\ \bibinfo {author} {\bibfnamefont {P.}~\bibnamefont
  {Saalfrank}},\ }\href@noop {} {\bibfield  {journal} {\bibinfo  {journal}
  {Phys. Chem. Chem. Phys.}\ }\textbf {\bibinfo {volume} {25}},\ \bibinfo
  {pages} {11771} (\bibinfo {year} {2023}{\natexlab{b}})}\BibitemShut {NoStop}%
\bibitem [{\citenamefont {Fischer}\ and\ \citenamefont
  {Saalfrank}(2021)}]{Fischer2021}%
  \BibitemOpen
  \bibfield  {author} {\bibinfo {author} {\bibfnamefont {E.~W.}\ \bibnamefont
  {Fischer}}\ and\ \bibinfo {author} {\bibfnamefont {P.}~\bibnamefont
  {Saalfrank}},\ }\href@noop {} {\bibfield  {journal} {\bibinfo  {journal} {J.
  Chem. Phys.}\ }\textbf {\bibinfo {volume} {154}},\ \bibinfo {pages} {104311}
  (\bibinfo {year} {2021})}\BibitemShut {NoStop}%
\bibitem [{\citenamefont {Kosloff}(1988)}]{Kosloff1988}%
  \BibitemOpen
  \bibfield  {author} {\bibinfo {author} {\bibfnamefont {R.}~\bibnamefont
  {Kosloff}},\ }\href@noop {} {\bibfield  {journal} {\bibinfo  {journal} {J.
  Phys. Chem.}\ }\textbf {\bibinfo {volume} {92}},\ \bibinfo {pages} {2087}
  (\bibinfo {year} {1988})}\BibitemShut {NoStop}%
\bibitem [{\citenamefont {Marston}\ and\ \citenamefont
  {Balint-Kurti}(1989)}]{Marston1989}%
  \BibitemOpen
  \bibfield  {author} {\bibinfo {author} {\bibfnamefont {C.~C.}\ \bibnamefont
  {Marston}}\ and\ \bibinfo {author} {\bibfnamefont {G.~G.}\ \bibnamefont
  {Balint-Kurti}},\ }\href@noop {} {\bibfield  {journal} {\bibinfo  {journal}
  {J. Chem. Phys.}\ }\textbf {\bibinfo {volume} {91}},\ \bibinfo {pages}
  {3571--3576} (\bibinfo {year} {1989})}\BibitemShut {NoStop}%
\bibitem [{\citenamefont {Sathyamurthy}\ and\ \citenamefont
  {Mahapatra}(2021)}]{sathyamurthy2021time}%
  \BibitemOpen
  \bibfield  {author} {\bibinfo {author} {\bibfnamefont {N.}~\bibnamefont
  {Sathyamurthy}}\ and\ \bibinfo {author} {\bibfnamefont {S.}~\bibnamefont
  {Mahapatra}},\ }\href@noop {} {\bibfield  {journal} {\bibinfo  {journal}
  {Phys. Chem. Chem. Phys.}\ }\textbf {\bibinfo {volume} {23}},\ \bibinfo
  {pages} {7586} (\bibinfo {year} {2021})}\BibitemShut {NoStop}%
\bibitem [{\citenamefont {Klamroth}(2003)}]{Klamroth2003}%
  \BibitemOpen
  \bibfield  {author} {\bibinfo {author} {\bibfnamefont {T.}~\bibnamefont
  {Klamroth}},\ }\href@noop {} {\bibfield  {journal} {\bibinfo  {journal}
  {Phys. Rev. B}\ }\textbf {\bibinfo {volume} {68}},\ \bibinfo {pages} {245421}
  (\bibinfo {year} {2003})}\BibitemShut {NoStop}%
\bibitem [{\citenamefont {Krause}, \citenamefont {Klamroth},\ and\
  \citenamefont {Saalfrank}(2005)}]{Krause2005}%
  \BibitemOpen
  \bibfield  {author} {\bibinfo {author} {\bibfnamefont {P.}~\bibnamefont
  {Krause}}, \bibinfo {author} {\bibfnamefont {T.}~\bibnamefont {Klamroth}},\
  and\ \bibinfo {author} {\bibfnamefont {P.}~\bibnamefont {Saalfrank}},\
  }\href@noop {} {\bibfield  {journal} {\bibinfo  {journal} {J. Chem. Phys.}\
  }\textbf {\bibinfo {volume} {123}},\ \bibinfo {pages} {074105} (\bibinfo
  {year} {2005})}\BibitemShut {NoStop}%
\bibitem [{\citenamefont {Rohringer}, \citenamefont {Gordon},\ and\
  \citenamefont {Santra}(2006)}]{Santra2006}%
  \BibitemOpen
  \bibfield  {author} {\bibinfo {author} {\bibfnamefont {N.}~\bibnamefont
  {Rohringer}}, \bibinfo {author} {\bibfnamefont {A.}~\bibnamefont {Gordon}},\
  and\ \bibinfo {author} {\bibfnamefont {R.}~\bibnamefont {Santra}},\
  }\href@noop {} {\bibfield  {journal} {\bibinfo  {journal} {Phys. Rev. A}\
  }\textbf {\bibinfo {volume} {74}},\ \bibinfo {pages} {043420} (\bibinfo
  {year} {2006})}\BibitemShut {NoStop}%
\bibitem [{\citenamefont {Klinkusch}, \citenamefont {Saalfrank},\ and\
  \citenamefont {Klamroth}(2009)}]{Klinkusch2009}%
  \BibitemOpen
  \bibfield  {author} {\bibinfo {author} {\bibfnamefont {S.}~\bibnamefont
  {Klinkusch}}, \bibinfo {author} {\bibfnamefont {P.}~\bibnamefont
  {Saalfrank}},\ and\ \bibinfo {author} {\bibfnamefont {T.}~\bibnamefont
  {Klamroth}},\ }\href@noop {} {\bibfield  {journal} {\bibinfo  {journal} {J.
  Chem. Phys.}\ }\textbf {\bibinfo {volume} {131}},\ \bibinfo {pages} {114304}
  (\bibinfo {year} {2009})}\BibitemShut {NoStop}%
\bibitem [{\citenamefont {Riso}\ \emph {et~al.}(2022)\citenamefont {Riso},
  \citenamefont {Haugland}, \citenamefont {Ronca},\ and\ \citenamefont
  {Koch}}]{Riso2022}%
  \BibitemOpen
  \bibfield  {author} {\bibinfo {author} {\bibfnamefont {R.~R.}\ \bibnamefont
  {Riso}}, \bibinfo {author} {\bibfnamefont {T.~S.}\ \bibnamefont {Haugland}},
  \bibinfo {author} {\bibfnamefont {E.}~\bibnamefont {Ronca}},\ and\ \bibinfo
  {author} {\bibfnamefont {H.}~\bibnamefont {Koch}},\ }\href@noop {} {\bibfield
   {journal} {\bibinfo  {journal} {Nat. commun.}\ }\textbf {\bibinfo {volume}
  {13}},\ \bibinfo {pages} {1368} (\bibinfo {year} {2022})}\BibitemShut
  {NoStop}%
\bibitem [{\citenamefont {Kaufmann}, \citenamefont {Baumeister},\ and\
  \citenamefont {Jungen}(1989)}]{Kaufmann1989}%
  \BibitemOpen
  \bibfield  {author} {\bibinfo {author} {\bibfnamefont {K.}~\bibnamefont
  {Kaufmann}}, \bibinfo {author} {\bibfnamefont {W.}~\bibnamefont
  {Baumeister}},\ and\ \bibinfo {author} {\bibfnamefont {M.}~\bibnamefont
  {Jungen}},\ }\href@noop {} {\bibfield  {journal} {\bibinfo  {journal} {J.
  Phys. B: At. Mol. Opt. Phys.}\ }\textbf {\bibinfo {volume} {22}},\ \bibinfo
  {pages} {2223} (\bibinfo {year} {1989})}\BibitemShut {NoStop}%
\bibitem [{\citenamefont {Szabo}\ and\ \citenamefont
  {Ostlund}(1996)}]{SzaboOstlund}%
  \BibitemOpen
  \bibfield  {author} {\bibinfo {author} {\bibfnamefont {A.}~\bibnamefont
  {Szabo}}\ and\ \bibinfo {author} {\bibfnamefont {N.}~\bibnamefont
  {Ostlund}},\ }\href@noop {} {\emph {\bibinfo {title} {Modern Quantum
  Chemistry: Introduction to Advanced Electronic Structure Theory}}}\ (\bibinfo
   {publisher} {Dover Publications, Inc.},\ \bibinfo {address} {Mineola},\
  \bibinfo {year} {1996})\BibitemShut {NoStop}%
\end{thebibliography}%
%
\end{document}